\documentclass[fleqn,usenatbib]{mnras}

\usepackage{newtxtext,newtxmath}

\usepackage[T1]{fontenc}

\DeclareRobustCommand{\VAN}[3]{#2}
\let\VANthebibliography\thebibliography
\def\thebibliography{\DeclareRobustCommand{\VAN}[3]{##3}\VANthebibliography}

\usepackage{graphicx,url}	
\usepackage{amsmath}	
\usepackage[english]{babel}
\usepackage{multicol}
\usepackage[font=normalsize]{caption}
\usepackage{relsize}
\usepackage{bm}
\usepackage[flushleft]{threeparttable}
\usepackage[none]{hyphenat}

\title[AGN dust models]{Towards an observationally motivated AGN dusty torus model. I. Dust chemical composition from the modeling of Spitzer spectra}

\author[Reyes-Amador et al.]{%
Omar Ulises Reyes-Amador,$^{1}$\thanks{E-mail: o.reyes@irya.unam.mx }
Jacopo Fritz,$^{1}$
Omaira González-Martín,$^{1}$
Sundar Srinivasan,$^{1}$
\newauthor
Maarten Baes,$^{2}$
Enrique Lopez-Rodriguez,$^{3}$
Natalia Osorio-Clavijo,$^{1}$
Cesar Iván Victoria-Ceballos,$^{1}$
\newauthor
Marko Stalevski,$^{2,4}$
C. Ramos Almeida, $^{5,6}$
\\
$^{1}$Instituto de Radioastronomía y Astrofísica (IRyA), Universidad Nacional Autónoma de México (UNAM), Antigua Carretera a Pátzcuaro, 8701, \\
Ex-Hda. San José de la Huerta, Morelia, Michoacán, 58089, México\\
$^{2}$Sterrenkundig Observatorium, Universiteit Gent, Krijgslaan 281 S9, 9000 Gent, Belgium \\
$^{3}$Kavli Institute for Particle Astrophysics \& Cosmology (KIPAC), Stanford University, Stanford, CA 94305, USA \\
$^{4}$Astronomical Observatory, Volgina 7, 11060 Belgrade, Serbia\\
$^{5}$Instituto de Astrofísica de Canarias, Calle Vía Láctea, s/n, 38205, La Laguna, Tenerife, Spain\\
$^{6}$Departamento de Astrofísica, Universidad de La Laguna, 38206, La Laguna, Tenerife, Spain
}

\date{Accepted XXX. Received YYY; in original form ZZZ}

\pubyear{2022}
\begin{document}
\label{firstpage}
\pagerange{\pageref{firstpage}--\pageref{lastpage}}
\maketitle

\begin{abstract}
Spectral energy distribution (SED) fitting is one of most commonly used techniques to study the dust properties in Active Galactic Nuclei (AGN).
Works implementing this technique commonly use radiative transfer models that assume a variety of dust properties. Despite the key role of this aspect, limited effort has been put forward to explore the chemical composition, the role of different optical properties, and the grain size distribution of dust, all of which can have a substantial impact on the theoretical radiative transfer calculations. In this work, 
we explore the role of the dust chemical composition in the AGN dusty torus through SED fitting to \emph{Spitzer}/IRS spectra of a sample of 49 nearby AGN with silicate features in emission. We implement a mineralogy model including the popular astronomical silicates and a set of oxides and amorphous silicates with different grain sizes. We find that best fits use principally porous alumina, periclase, and olivine. In terms of mass fractions, $\sim99\%$ of the dust is composed of dust grains of size $\rm{0.1 \mu m}$, with a $<1\%$ contribution from $\rm{3 \mu m}$ grains. Moreover, the astronomical silicates have a very low occurrence in the best fits, suggesting that they are not the most suited dust species to reproduce the silicate features in our sample. 
\end{abstract}




\section{Introduction} \label{sec:intro}

The ``Unified Model'' (or Unification Scheme) proposed by \citet{Antonucci1993} \citep[see also][]{Urry1995} is the most accepted model that can explain the classification of Active Galactic Nuclei (AGN) into type 1 and type 2. According to this scheme, the centres of all AGN host a supermassive black hole (SMBH) surrounded by an accretion disc that is heated up to $\sim$10$^5$ K, and a region of clouds of gas moving at velocities of $\sim$10$^4$ km s$^{-1}$, commonly called the broad-line region (BLR). The last two components are responsible for producing the strong ultraviolet (UV)/optical emission and the broad emission lines, respectively. At distances from the centre where the temperature is lower than $\sim$10$^3$ K, dust can coagulate \citep{Maiolino2001} surviving the intense radiation field. It is presumed to be distributed in a geometrically and optically thick obscuring structure, known as the ``dusty torus'', which surrounds the AGN \citep[see][for a recent review]{Ramos-Almeida2017}. 

This structure is a cornerstone of the Unification Scheme, providing a simple explanation of the diversity of features observed in type 1 and type 2 AGN, whose key is the viewing angle at which the torus is observed. An issue that this structure helps to explain is the observation of polarized broad emission lines produced by the scattering on electrons and dust in type 2 objects \citep{Antonucci1985,Moran2000,Nagao2004,Moran2007,Trump2011,Pan2019}. Furthermore, the dusty torus heats up by absorbing the radiation coming from the accretion disc and re-emits it at infrared (IR) wavelengths as thermal radiation producing the "infrared bump", a strong IR continuum emission observed from $\rm{\sim1}$ to $\rm{\sim50\ \mu m}$ in the spectral energy distribution (SED) of AGN. This is why observations in the Near-IR (NIR) to Mid-IR (MIR) offer a window that is crucial to study the AGN dusty torus. The NIR continuum emission has been associated with thermal dust emission ranging from 1000 to 1800 K \citep{Rees1969}. \citet{Barvainis1987} has modeled this emission using graphite dust grains at their sublimation temperature of $\sim1500$ K. Also, spectral features at $\rm{\sim10}$ and $\rm{\sim18\ \mu m}$ have been observed \citep{Rieke1975,Kleinmann1976} such as in the interstellar medium (ISM) of our Galaxy \citep{Stein1969,Hackwell1970}. These features are attributed to the Si-O stretching and O-Si-O bending modes of amorphous silicates \citep{Smith2010}, and provide interesting clues to the characteristics of dust. 


Since the 1990s, efforts in modeling the infrared SED of AGN produced by the dusty torus (hereinafter, dusty torus models) have been made using different methods and characteristics. 
One of the methods that has allowed to derive physical properties and improve the understanding of the dusty torus is the modeling of the MIR emission using radiative transfer (RT). Most of the models consider a flared disc as an approximation of the torus geometry while the density distribution of the dust has been modeled assuming a homogeneous (smooth) dust distribution \citep{GranatoDanese1994,Efstathiou1995,vanBemmel2003,Schartmann2005,Fritz2006}, a clumpy dust distribution (dust arranged in clouds) \citep{Nenkova2002,Nenkova2008b,Dullemond2005,Honig2006,Honig2010,Honig2013,Honig&Kishimoto2010,Honig&Kishimoto2017}, and a combination of clumpy and smooth distributions called ``two-phase medium'' \citep{Stalevski2012,Stalevski2016,Siebenmorgen2015,Gonzalez-Martin2023}. In addition, a model with an outflowing polar wind of dust and a geometrically thin disc, known as the ``disc+wind'' model, has been proposed as a new alternative of dusty torus models \citep{Honig&Kishimoto2017,Stalevski2017,Stalevski2019}. 

A poorly explored aspect of the dusty torus models is the dust chemical composition. Most of the models mentioned above assume that the dust primarily consists of the standard composition of the ISM: separate populations of graphite and silicate dust grains \citep{Mathis1977,DraineLi1984}. The silicate spectral features are typically modeled using the optical and calorimetric properties of ``astronomical silicates'' \citep{DraineLi1984,LaorDraine1993,WingartnerDraine2001,Draine2003,Min2007,DraineLi2007} because they have been able to reproduce the observed characteristics in the MIR emission of galaxies. 
However, the MIR emission of some AGN shows atypical characteristics, e.g., the peak of the $\rm{\sim10\ \mu m}$ silicate feature is shifted to longer wavelengths, the shapes of both silicate features are broader and the relative intensity of the feature at $\rm{\sim18\ \mu m}$ is larger than the one at $\rm{\sim10\ \mu m}$ \citep[e.g.][]{Sturm2005, Li2008, Smith2010}. These atypical characteristics can not be well reproduced by the standard composition of the ISM nor by the radiative transfer models mentioned above, suggesting that there are differences between the AGN dust and the ISM in the composition, size distribution and degree of crystallization of the silicate dust \citep{Sturm2005}. Therefore, alternative chemical compositions for the dust could help to explain them. 

Within the works that have investigated alternative chemical compositions for the AGN dust, \citet{Markwick-Kemper2007} and \citet{Srinivasan2017} followed a particular approach. Both teams have applied the SED fitting technique adopting a mineralogy model composed of a power-law (PL) continuum and dust species such as porous alumina, periclase, olivine, Mg-rich olivine, and forsterite to study their contribution to the silicate emission features in the MIR spectrum of AGNs. The main differences between these two analyses are that \citet{Markwick-Kemper2007} studied only the BAL quasar PG~ 2112+059 and considered spherical grains with a fixed grain size and non-spherical grains with a fixed volume, while \citet{Srinivasan2017} focused on a sample of QSOs using dust modeled as a continuous distribution of ellipsoids (CDE) with a fixed volume and also include clinoenstatite in their set of dust species. 

The scientific community is currently dealing with a rich set of dusty torus models, each of them created with a different radiative transfer code and assuming fairly different ingredients: the accretion disc SED, the chemical composition, optical properties, grain sizes, and density distribution of the dust. Yet, these models are only characterized by this latter aspect, biasedly considered as the one mainly defining their characteristics. These sets of models have been used to derive the geometrical and density distribution of dust, by fitting NIR and MIR photometric data, spectra, or a combination of the two \citep{Hatziminaoglou2008,Ramos-Almeida2009,Ramos-Almeida2011,Berta2013,Martinez-Paredes2017,Martinez-Paredes2020,Martinez-Paredes2021,Gonzalez-Martin2019II,Gonzalez-Martin2023,Boquien2019}. However, the efforts of these dusty torus models to obtain consistent results among them when compared with observations have been strongly linked only to the dust density distribution, so it is important to also investigate the other parameters already mentioned. 

Here, we study the chemical composition of AGN dust in order to go beyond the ``assumption by default'' of this parameter in the dusty torus models. 
In this paper, we investigate the existence of dust models better suited to reproduce the characteristics of typical AGN MIR spectra, with a particular focus on the silicate features, hence physically motivating the choice of the dust chemical composition to adopt in the production of future radiative transfer models. The manuscript is organized as follows: Section 2 shows the sample and data. The methodology is presented in Section 3. Results are included in Section 4. Finally, Section 5 shows a discussion of the results, while conclusions are presented in Section 6.

\section{Observational data collection} \label{sec:data_collection}

We rely on the sample of nearby ($z \lid 0.1$) AGN studied by \citet{Gonzalez-Martin2023}, which is composed by 68 AGN with available \emph{Spitzer}/IRS spectra, good signal-to-noise ratio, and dominated by the AGN dust at the range of the \emph{Spitzer}/IRS spectra (5-35~$\rm{\mu m}$). However, data above 32 $\mu$m were not taken into account for our analysis because the signal-to-noise at those wavelengths is significantly lower. We ensure that the spectra are dominated by AGN dust emission (at least for 80\% of the flux in the range 5-30~$\rm{\mu m}$) because a detailed spectral decomposition was done by \citet{Gonzalez-Martin2019I} and references there in, using ISM, stellar and AGN-heated dust components. The sample includes 43 type 1 and 25 type 2 AGN and covers the X-ray luminosity range from low-luminosity ($L_{\text{X}} \sim 10^{41}\ {\text{erg~s}}^{-1}$,  e.g., M\,106), to intermediate-luminosity ($L_{\text{X}} \sim 10^{43.5}\ {\text{erg~s}}^{-1}$, e.g, NGC~3783), and high-luminosity AGN ($L_{\text{X}} \sim 10^{44.5}\ {\text{erg~s}}^{-1}$, e.g., PG~0804+761).  

\begin{table*}
\begin{threeparttable}
\small 
\renewcommand{\arraystretch}{0.9}
\begin{center}
\caption{\normalsize Relevant information of the sample.}
\label{tab:sample}
\begin{tabular}{lcccclcccc}\hline \hline
Object name        &  $z$     &Class& $\rm{log(L_{X})}$ & $\rm{S_{10\mu m}}$ & Object name        &  $z$     &Class& $\rm{log(L_{X})}$ & $\rm{S_{10\mu m}}$ \\ 
(1) & (2) & (3) & (4) & (5) & (6) & (7) & (8) & (9) & (10) \\ \hline 
I~Zw~1                     &  0.059 & S1  &  43.85 & -0.291$\pm$0.008 & 
FAIRALL~9                 &  0.047 & S1  &  44.09  & -0.184$\pm$0.004 \\
NGC~526A                  &  0.019 & S1  &  43.78 & -0.136$\pm$0.004 & 
Mrk~1018                  &  0.042 & S1  &  42.82 & -0.260$\pm$0.002 \\ 
Mrk~590                   &  0.021 & S1  &  43.23  & -0.200$\pm$0.004 &
NGC~1052                  &  0.005 & S1  &  42.24  & -0.265$\pm$0.002 \\ 
NGC~1275                  &  0.016 & S2  &  43.76 & -0.393$\pm$0.008 &
ESO~548-G081              &  0.014 & S1  &  43.29  & -0.271$\pm$0.003 \\ 
3C~120                    &  0.033 & S1  &  44.38  & -0.305$\pm$0.004 & 
MCG~-01-13-025            &  0.016 & S1  &  43.29 & -0.575$\pm$0.001\\ 
CGCG~420-015              &  0.029 & S2  &  43.74  & -0.138$\pm$0.005 &
Ark~120                   &  0.032 & S1  &  43.78  & -0.266$\pm$0.006 \\
PICTOR~A                  &  0.035 & S1  &  44.03 & -0.502$\pm$0.002 & 
Mrk~1210                  &  0.013 & S1  &  43.37  & -0.191$\pm$0.010 \\
PG~0804+761               &  0.100 & S1  &  44.46 & -0.433$\pm$0.002 & 
MCG~+04-22-042            &  0.032 & S1  &  43.73 & -0.207$\pm$0.005 \\ 
Mrk~110                   &  0.033 & S1  &  44.25 &-0.243$\pm$0.002 & 
Mrk~705                   &  0.029 & S1  &  43.41  & -0.112$\pm$0.004 \\
ESO~374-G044              &  0.028 & S2  &  43.64 & -0.100$\pm$0.002 & 
Mrk~417                   &  0.033 & S2  &  43.91  & -0.107$\pm$0.004 \\
NGC~3783                  &  0.011 & S1  &  43.56 & -0.258$\pm$0.010 & 
UGC~6728                  &  0.007 & S1  &  42.40  & -0.223$\pm$0.002 \\
2MASX~J11454045-1827149   &  0.033 & S1  &  44.08 & -0.223$\pm$0.003 & 
Ark~347                   &  0.022 & S2  &  43.52  & -0.095$\pm$0.027 \\
NGC~4151                  &  0.002 & S1  &  43.17 & -0.092$\pm$0.156 & 
PG~1211+143               &  0.090 & S1  &  43.70  & -0.315$\pm$0.003 \\
M~106                     &  0.002 & S1  &  41.06 & -0.394$\pm$0.003 & 
NGC~4507                  &  0.012 & S2  &  43.76 & -0.193$\pm$0.015 \\ 
NGC~4939                  &  0.009 & S2  &  42.81  & -0.243$\pm$0.001 &
II~SZ~010                  &  0.034 & S1  &  43.52 & -0.177$\pm$0.002 \\ 
MCG~-06-30-015            &  0.008 & S1  &  42.82 & -0.112$\pm$0.009 & 
IC~4329A                  &  0.016 & S1  &  43.77  & -0.116$\pm$0.048 \\
UM~614                    &  0.033 & S1  &  41.74 & -0.261$\pm$0.001 & 
Mrk~279                   &  0.030 & S1  &  43.87  & -0.108$\pm$0.006 \\
PG~1351+640               &  0.088 & S1  &  43.10  & -0.854$\pm$0.002 & 
NGC~5548                  &  0.025 & S1  &  43.76  & -0.212$\pm$0.005 \\
ESO~511-G030              &  0.015 & S1  &  43.65 & -0.365$\pm$0.002 & 
PG~1448+273               &  0.065 & S1  &  43.30  & -0.140$\pm$0.002 \\
Mrk~841                   &  0.036 & S1  &  44.01  & -0.101$\pm$0.005 &
Mrk~1392                  &  0.036 & S1  &  43.74  & -0.198$\pm$0.003\\ 
Mrk~1393                  &  0.054 & S1  &  43.80  & -0.118$\pm$0.001 &
Mrk~290                   &  0.030 & S1  &  43.68 & -0.225$\pm$0.002 \\ 
ESO~138-G001              &  0.009 & S2  &  42.55  & -0.147$\pm$0.161 &
Fairall~51                &  0.011 & S1  &  43.22  & -0.277$\pm$0.007 \\
ESO~141-G055              &  0.037 & S1  &  44.25 & -0.344$\pm$0.003 & 
NGC~6814                  &  0.003 & S1  &  42.59 & -0.203$\pm$0.008 \\ 
II~Zw~136                  &  0.078 & S1  &  43.50 & -0.068$\pm$0.004 & 
NGC~7213                  &  0.005 & S1  &  42.46  & -0.659$\pm$0.003 \\
PG~2304+042               &  0.042 & S1  &  43.40 & -0.569$\pm$0.001 &
& & & & \\
\hline\hline
\end{tabular}
\begin{tablenotes}
\item \noindent \textbf{Notes:} Columns 1 and 6 are the names of the objects. Columns 2 and 7 show the redshift. Columns 3 and 8 are the AGN classification: Seyfert 1 (S1) and Seyfert 2 (S2). Columns 4 and 9 are the logarithm of the X-ray 2-10 keV intrinsic luminosity in erg s$^{-1}$. Columns 5 and 10 show the $\rm{10\ \mu m}$ silicate strength. This information has been taken from \citet{Gonzalez-Martin2023}.
\end{tablenotes}
\end{center}
\end{threeparttable}
\end{table*}

\begin{figure}
    \centering
    \includegraphics[scale=0.52]{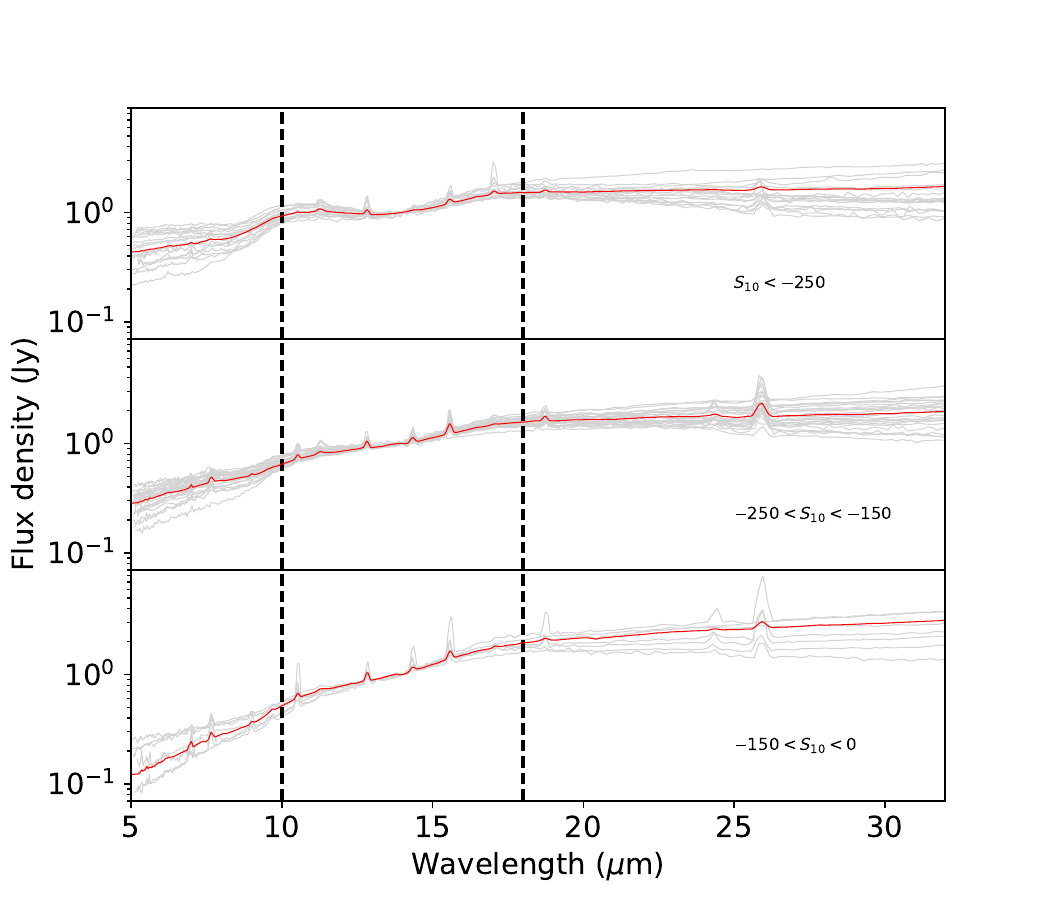}
    \caption{\emph{Spitzer}/IRS spectra of our sample of 49 objects are here shown in light-gray color. The red lines represent the average of the spectra in each panel. Spectra with the strongest to the weakest  silicate feature at $\rm{10\ \mu m}$ are shown from top panel to the bottom. The $\rm{\rm{S_{10\mu m}}}$ bins that define in which panel each spectrum lands are $S_{10}<-250$, $-250<S_{10}<-150$, and $-150<S_{10}<0$, respectively. All spectra are normalized to the flux at $\rm{14\ \mu m}$. The black dashed vertical lines correspond to 10 and $\rm{18\ \mu m}$, respectively. }
    \label{fig:all_spectra}
\end{figure}

Because the model we adopt to reproduce these spectra assumes that this emission arises from an optically thin (at these wavelengths) dust layer (see Section~\ref{sec:fitting_model} for details), in this work we only consider objects showing silicate emission features in their spectra. To discern those with silicate features in emission from those in absorption, the $\rm{10\ \mu m}$ silicate strength (denoted as $\rm{\rm{S_{10\mu m}}}$) was adopted:
\begin{equation} \label{eq:silicates_strength}
    \text{S}_{\lambda} = -\ln\left(\frac{F_{\nu}(\lambda)}{F_{\nu}(\text{continuum})}\right).
\end{equation}
The latter uses the sign convention for which absorption features have positive values and emission features have negative ones. It is worth mentioning that although we use $\rm{10\ \mu m}$ to refer to this silicate feature and its strength, the calculation of $\rm{\rm{S_{10\mu m}}}$ was carried out taking the flux at $\rm{\lambda = 9.7\ \mu m}$. This parameter only was used to select those spectra with this silicate feature in emission. From the 68 objects of the complete sample, 49 objects show $\rm{S_{10\mu m}}$ in emission, of which 41 are type 1 and 8 are type 2. Table~\ref{tab:sample} shows the redshift ($z$), the AGN classification, the X-ray 2-10 keV intrinsic luminosity, and the 10 $\mu$m silicate strength
for the 49 objects of our sample. Their spectra are shown in Figure~\ref{fig:all_spectra}. The spectra are displayed at rest-frame wavelengths, and the analysis that follows is performed on redshift-corrected data. We note that, in most of the spectra, the peak of the $\rm{10\ \mu m}$ silicate emission feature is slightly shifted to longer wavelengths with respect to the nominal $\rm{9.7\ \mu m}$ value, similar to NGC~3998 \citep{Sturm2005} and M\,81 \citep{Smith2010}, while the peak of the $\rm{18\ \mu m}$ feature is slightly shifted to shorter wavelengths. Both characteristics are reported by \citet{Martinez-Paredes2020} as well. 

\section{Data analysis and methodology} \label{sec:methodology}

We explore the chemical composition of the dust grains of dusty tori through the SED fitting method using a model for the MIR spectra considering a set of different combinations of dust species. All the models are statistically tested to obtain the best-fit combination of the parameters and to assess their goodness-of-fit. In the following sections, we explain the data processing (mainly the treatment to exclude emission lines), the details of the fitting model, the procedure to obtain the optical properties of the dust species to be implemented in the model, the set of fitted combinations, and the criteria to evaluate them.

\subsection{Masking emission lines}

It is important to mention that we do not consider emission lines produced by ionized gas because our analysis is focused on the AGN dust continuum emission only. 
We have explored spectral decomposition using pahfitMCMC, a modified version of PAHFIT \citep{Smith2007} that includes silicate features in emission which has been used by \citet{Garcia-Bernete2022} and references therein, in order to obtain emission-lines and PAHs subtracted spectra. We have found that pahfitMCMC overestimates the emission of PAHs against the emission of silicates, making that the $\rm{18\ \mu m}$ silicate feature disappears. Furthermore, this technique is not convenient for our analysis because pahfitMCMC assumes the cold silicates of \citet{Ossenkopf1992} to fit the silicate features in emission which, being closer to the characteristics of the ISM dust, is not the most appropriate choice for dust in AGN. This kind of approach makes the PAHFIT decomposition model dependent.

We have removed any emission line located in the 5 to 32 $\mu$m range for each of the 49 spectra, so that both the continuum and the silicate feature emission could be properly measured, thus avoiding possible biases due to their presence. This was done by first carefully inspecting each spectrum to identify the presence of any prominent emission line and then masking the corresponding spectral region by ignoring the flux at those wavelengths taking good care to check that the wings were masked as well. The removed emission lines are listed in Table~\ref{tab:masked_emission_lines} and correspond to the most prominent lines in Figure~\ref{fig:all_spectra}.

\begin{table}
\begin{threeparttable}
\small 
\renewcommand{\arraystretch}{1.2}
\begin{center}
\caption{\normalsize Emission lines masked in the spectra.}
\begin{tabular}{lcc}\hline \hline
Emission line & $\lambda_{\text{peak}}$ & $\lambda_{\text{range}}=[\lambda_1,\lambda_2]$ \\ 
& ($\mu$m) & ($\mu$m) \\ 
(1) & (2) & (3) \\ \hline
Unidentified & $\sim5.40$ & [5.17,5.67] \\
PAH & $\sim6.20$ & [5.82,6.50] \\
$[\text{Ar II}]$ & $\sim$7.00 & [6.75,7.05] \\
$[\text{Ne IV}]$ & $\sim$7.65 & [7.16,8.10] \\
$[\text{Ar III}]$ & $\sim$9.00 & [8.84,9.16] \\
H2(S3) & $\sim$9.72 & [9.55,9.84] \\
$[\text{S IV}]$ & $\sim$10.55 & [10.23,10.70] \\
PAH & $\sim11.30$ & [11.07,11.52] \\
$[\text{Ne II}]$ & $\sim$12.85 & [12.53,12.95] \\
$[\text{Ne V}]$ & $\sim$14.35 & [14.00,14.55] \\
$[\text{Ne III}]$ & $\sim$15.57 & [15.22,15.85] \\
H2(S1) & $\sim$17.02 & [16.81,17.48] \\
$[\text{Ne V}]$ & $\sim$24.36 & [23.80,24.75] \\
$[\text{O IV}]$ & $\sim$25.95 & [25.25,26.35] \\
\hline\hline
\end{tabular}
\label{tab:masked_emission_lines}
\begin{tablenotes}
\item \noindent \textbf{Notes:} Column 1 is the identifier of the emission line. Columns 2 and 3 are the wavelength peak and range of the emission line, respectively.
\end{tablenotes}
\end{center}
\end{threeparttable}
\end{table}

\subsection{The fitting model} \label{sec:fitting_model}

We use the mineralogy model reported by \citet{Markwick-Kemper2007} and \citet{Srinivasan2017} to fit the spectra, as the sum of two components: a thermal continuum emission, plus the silicate resonant features accounted for by the extinction efficiencies of the various silicate species. This model is a linear combination of the emission features of a number $N$ of dust species and a power-law continuum ($F_{\lambda,\text{c}}$): 
\begin{equation} 
\label{eq:minerology_model}
F_{\lambda,\text{mod}} 
=
F_{\lambda, {\rm c}}
\left(1+\sum_{j=1}^{N} c_j Q_{\text{ext,cs}}^{\lambda,j}\right),
\end{equation}
where $Q_{\text{ext,cs}}^{\lambda,j}$ and $c_j$ are the continuum-subtracted extinction efficiency and the relative number of dust grains (or mass fraction) of the $j^{\text{th}}$ dust species, respectively. 



\subsection{Calculation of the $Q_{\text{ext,cs}}$} 
\label{sec:Calculation_Qext,cs}

\begin{figure}
    \centering
    \includegraphics[scale=0.55]{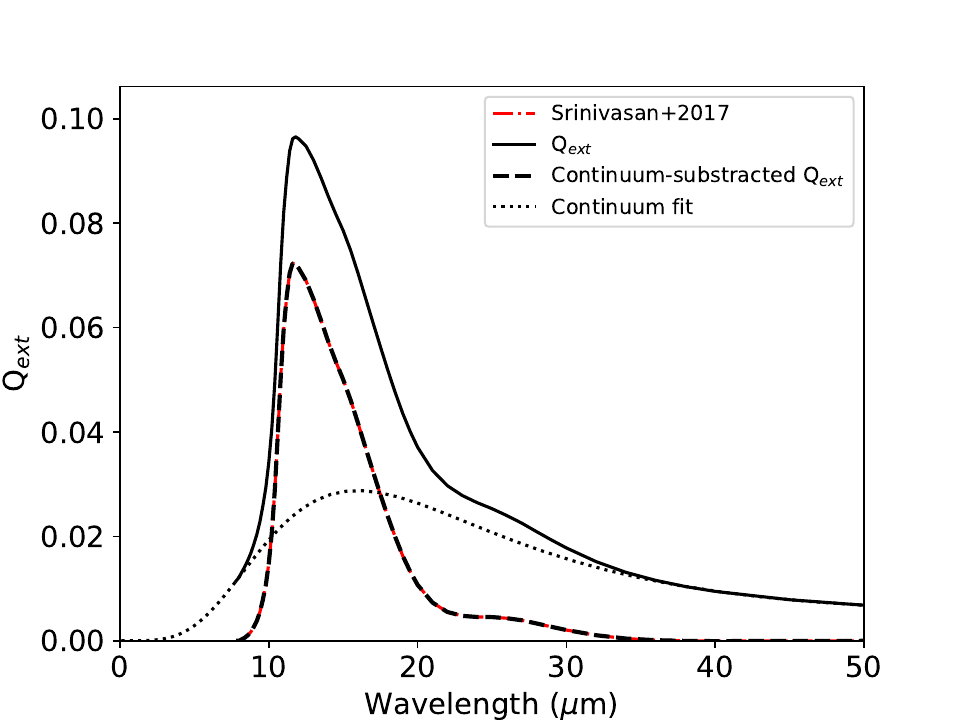}
    \caption{Extinction efficiency for porous alumina. The solid line is our calculated extinction efficiency $Q_{\text{ext}}^{\lambda}$. The dotted line is a robust quintic polynomial fit representing the $Q_{\text{cont}}^{\lambda}$ that we used to compute $Q_{\text{ext,cs}}^{\lambda}$ (black dashed line). Red dash-dotted line is the $Q_{\text{ext,cs}}^{\lambda}$ calculated by \citet{Srinivasan2017}.}
    \label{fig:Qext_porous_alumina}
\end{figure}

We calculate the continuum-subtracted extinction efficiencies $Q_{\text{ext,cs}}^{\lambda}$, using a similar procedure as the one described in  \citet{Srinivasan2017}. First, we define $Q_{\text{ext}}^{\lambda}$ in the 4-8 $\mu$m and 35-50 $\mu$m wavelength ranges as the ``continuum'' $Q_{\text{cont}}^{\lambda}$. We subsequently fit a quadratic, a cubic, or robust quintic polynomial to these ranges in order to guess the continuum under the feature in the 8-35 $\mu$m range and to obtain $Q_{\text{ext,cs}}^{\lambda}$ by
\begin{equation} 
\label{eq:continuum-subtracted}
Q_{\text{ext,cs}}^{\lambda} = Q_{\text{ext}}^{\lambda} - Q_{\text{cont}}^{\lambda} .
\end{equation}
The continuum subtraction procedure is performed for the extinction efficiencies $Q_{\text{ext}}^{\lambda}$ of all the species used in this work. As an example, in Figure~\ref{fig:Qext_porous_alumina} we show $Q_{\text{ext,cs}}^{\lambda}$ for porous alumina in comparison with the one from \citet{Srinivasan2017}, confirming that our procedure is consistent with theirs.  

\subsection{Search for the best continuum model} 
\label{sec:continuum}

In order to perform an optimal continuum subtraction from the MIR spectra, in addition to the power-law function from Eq.~(\ref{eq:minerology_model}), we also explore a cubic function (poly3) defined in logarithmic space to fit the continuum similarly to \citet{Spoon2007,Sirocky2008,MendozaCastrejon2015}. As the power law is a first-degree polynomial in logarithmic space:
\begin{equation}
    f(x) = p_1x^{p_2} \iff \log f(x) = \log p_1 +p_2\log x ,
\end{equation} 
we add two terms to make it a poly3: 
\begin{equation}
    \log f(x) = \log p_1 +p_2\log x + p_3\, (\log x)^2 +p_4\, (\log x)^3,
\end{equation} 
To optimize the modeling and check for possible biases, we repeat the fitting procedure of the 49 objects using these two different functions for the continuum, obtaining two fit models for each object, one with power law and other with poly3. As these models have differing numbers of parameters, we compare the qualities of the resulting fits using the Akaike Information Criterion \citep[AIC;][see Appendix~\ref{append:statistical_procedure}]{Akaike1974}. We find that, although poly3 gives an AIC value lower than power law, the difference between these values is $\Delta$AIC$<2$, meaning that the difference in quality between the fits obtained with the power law and poly3 functions is not statistically significant. In other words, the continuum subtraction is not sensitive to the specific choice of the functional form.
Hence, for the fits that we will present in the remainder of this paper, we use poly3 to model the continuum since the AIC values show this is the preferred model.  

\subsection{Dust species to model the silicate features} 
\label{sec:dust_species}

We explore as many compositions as possible, including dust species adopted in the pure phenomenological analysis and those used in the RT dusty models mentioned in Section~\ref{sec:intro}, to check their validity in reproducing the MIR emission of our sample. RT models generally assume a standard dust composition as the one found in the ISM of the Galaxy, which includes the so-called astronomical silicates that have been used to explain the silicate features at 9.7 and $\rm{18 \mu m}$ observed in spectra of ISM scenarios such as circumstellar dust and nebulae. In order to provide a comprehensive comparison between different dust properties, we include also astronomical silicates from \citet{DraineLi2007} (DLSil) in our analysis. This adopts the same properties as \citet{LaorDraine1993} which, in turn, uses the same optical constants for graphite and silicates as the one calculated by \citet{DraineLi1984} and an extension into the X-ray domain. In addition, \citet{DraineLi2007} includes polycyclic aromatic hydrocarbon (PAH). We explore the astronomical silicates of \citet{LaorDraine1993} and, as expected, we do not find differences in the fits with respect to DLSil. We also include the astronomical silicates from \citet{Min2007} (MinSil), which considers both amorphous and crystalline silicates with a non-spherical shape, while DLSil considers only amorphous silicates with a spherical shape. 

\begin{table*}
\begin{threeparttable}
\small 
\renewcommand{\arraystretch}{1.2}
\begin{center}
\caption{\normalsize Chemical and physical information about dust species explored in this study.}
\begin{tabular}{lccccccc}\hline \hline
Dust species &  Formula & Dust species & Wavelength range & Grain density & \multicolumn{2}{c}{Grain sizes} & Optical constants \\
 label & & name & ($\mu$m) & (g cm$^{-3}$) & CDE & Mie & Reference \\ 
 (1) & (2) & (3) & (4) & (5) & (6) & (7) & (8) \\ \hline
PoAl & Al$_2$O$_3$ & Porous Alumina & 0.04 - 500 & 4.02 & 0.1 & - & \citet{Begemann1997} \\
Per & MgO & Periclase & 0.2 - 3333 & 3.56 & 0.1 & - & \citet{Hofmeister2003}\\
Ol & MgFeSiO$_4$ & Olivine & 0.2 - 500 & 3.79 & 0.1-3.0\tnote{b} &  0.1-3.0\tnote{b} & \citet{Dorschner1995}\\
MgOl & Mg$_{2}$SiO$_4$ & Mg-rich Olivine & 0.19 - 948 & 3.22 & 0.1 & - & \citet{Jaeger2003}\\
Forsterite\tnote{a} & Mg$_2$SiO$_4$ & Forsterite & 0.2 - 852 & 3.2 & 0.1 & - & \citet{Jaeger1998} \\
Enstatite\tnote{a} & MgSiO$_3$ & Clinoenstatite & 0.04 - 98 & 3.28 & 0.1 & - & \citet{Jaeger1998} \\
DLSil & - & Draine \& Li Silicates & 0.04 - 10$^5$ & 3.0 & - & 0.1-4.0\tnote{b} & \citet{DraineLi2007} \\
MinSil & - & Min Silicates & 10$^{-3}$ - 10$^3$ & 3.09 & 0.1-4.0\tnote{b} & - & \citet{Min2007} \\
\hline\hline
\end{tabular}
\label{tab:chemical_compositions}
\begin{tablenotes}
\item \noindent \textbf{Notes:} Columns 6 and 7 are the available grain sizes used in this study. (6) CDE: ellipsoidal grains with different sizes but having the characteristic radius at which the volume of a spherical grain is equal to the volume of an ellipsoidal grain. (7) Mie theory: radius of a spherical grain. Although MinSil grain sizes are in the CDE column, they are defined as non-spherical grains.
\item [a] These dust species are not considered in the fitting procedure.
\item [b] In these ranges, the used grain sizes are: for Olivine, 0.1, 1.0, 2.0, and 3.0 $\mu$m, for DLSil, 0.15, 1.19, 2.26, 3.46 and 4.27 $\mu$m; and for MinSil, 0.1, 1.0, 1.99, 3.16 and 3.98 $\mu$m.
\end{tablenotes}
\end{center}
\end{threeparttable}
\end{table*}

\begin{figure}
    \centering
    \includegraphics[scale=0.45]{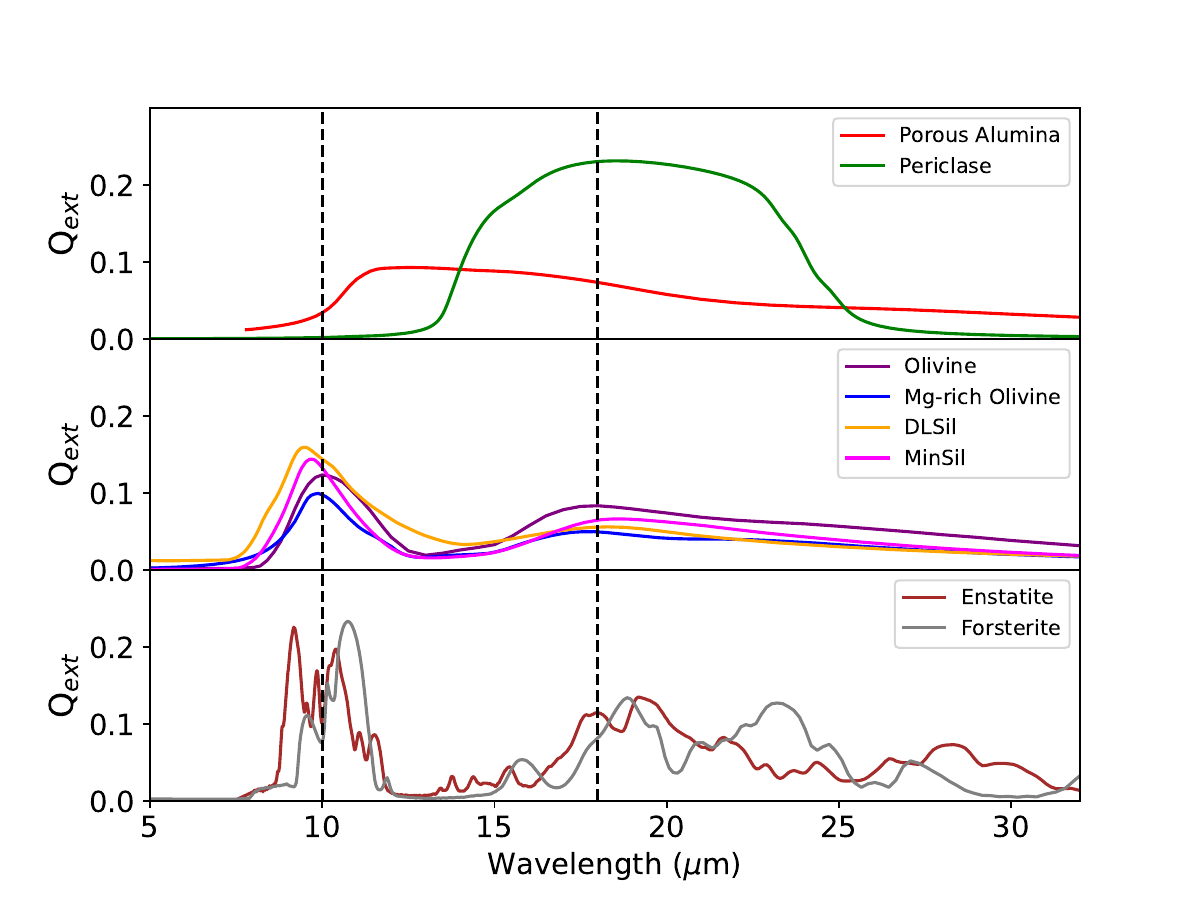}
    \caption{Extinction efficiencies of the dust species used in this work to fit the silicate features of the MIR spectra of the sample. The top panel shows the oxides porous alumina (red) and periclase (green). The middle panel shows the amorphous silicates olivine (purple), Mg-rich olivine (blue), DLSil (orange), and MinSil (magenta). The bottom panel shows the crystalline silicates Enstatite (brown) and Forsterite (gray). For all dust species in this plot, the $Q_{\text{ext}}$ corresponds to a grain size of $\rm{0.1\ \mu m}$. The black dashed vertical lines are at 10 and $\rm{18\ \mu m}$.}
    \label{fig:Qext_all_compositions}
\end{figure}

We also explore the six dust species used by \citet{Srinivasan2017} for their phenomenological SED fitting. We use their computed $Q_{\text{ext,cs}}^{\lambda}$ of the oxides porous alumina (corundum) and periclase, the amorphous silicates olivine and magnesium-rich (Mg-rich) olivine, and the crystalline silicates forsterite and clinoenstatite. Each of these is assumed to consist of a CDE of fixed volume, corresponding to a spherical grain with a radius of $\rm{0.1\ \mu m}$. Figure~\ref{fig:Qext_all_compositions} shows $Q_{\text{ext}}^{\lambda}$ as a function of  wavelength for all dust species explored in this work. It is divided into three panels depending on the type of material. The top panel shows the oxides, where it is possible to see that porous alumina has its major contribution between 10 and 20 $\mu$m and periclase contributes at wavelengths between 13 and 25 $\mu$m. The middle panel shows the amorphous silicates, which have two features around 10 $\mu$m and 18 $\mu$m, respectively. It is important to note that the peak wavelength of these features is slightly different for each dust species, e.g., olivine has its peak at 10 $\mu$m while the others have it at shorter wavelengths. Finally, the bottom panel shows the crystalline silicates, which show a large number of features at different wavelengths. 

Before proceeding to test the ability of the various dust grain combinations (as mentioned in Section~\ref{sec:sets_fit_combinations}) in reproducing a spectrum, we tested the six dust species from \citet{Srinivasan2017} as a single combination in each spectrum of the sample.
One very clear result was that clinoenstatite and forsterite have a negligible contribution in all cases, so we proceeded to carry out the fits considering only porous alumina, periclase, olivine, Mg-rich olivine, DLSil, and MinSil (see Table~\ref{tab:chemical_compositions}). For olivine, in addition to the $Q_{\text{ext,cs}}^{\lambda}$ for the CDE grain size of $\rm{0.1\ \mu m}$, we calculated and used the $Q_{\text{ext,cs}}^{\lambda}$ for CDE grain sizes of 1, 2, and 4~$\mu$m, and Mie (spherical particles) grain sizes of 0.1, 1, 2, and 3~$\mu$m. For both DLSil and MinSil we used $Q_{\text{ext}}^{\lambda}$ calculated for grain sizes of $\sim 0.1$, $\sim 1$, $\sim 2$, $\sim 3$, and $\sim 4~\mu$m (see notes in Table~\ref{tab:chemical_compositions}). Then we calculated $Q_{\text{ext,cs}}^{\lambda}$ for each grain size using Equation~(\ref{eq:continuum-subtracted}). A caveat of this work is that we use discrete grain sizes because, for those grain sizes, we found the $Q_{\text{ext}}^{\lambda}$ values in the literature (see Section~\ref{sec:caveats}). The refractive indices (or optical constants) $n(\lambda)$ and $k(\lambda)$ obtained from laboratory measurements are necessary to calculate $Q_{\text{ext}}^{\lambda}$, but this is out of the scope of this work. Moreover, in this analysis, we do not include the silicates from \citet{Ossenkopf1992}, which are used in the clumpy torus model from \citet{Nenkova2008a,Nenkova2008b}. Nevertheless, they are similar to the silicate species studied in our analysis and within the Qext range covered by them, and we have verified that they do not contribute to obtain better results.

\begin{figure}
    \centering
    \includegraphics[scale=0.45]{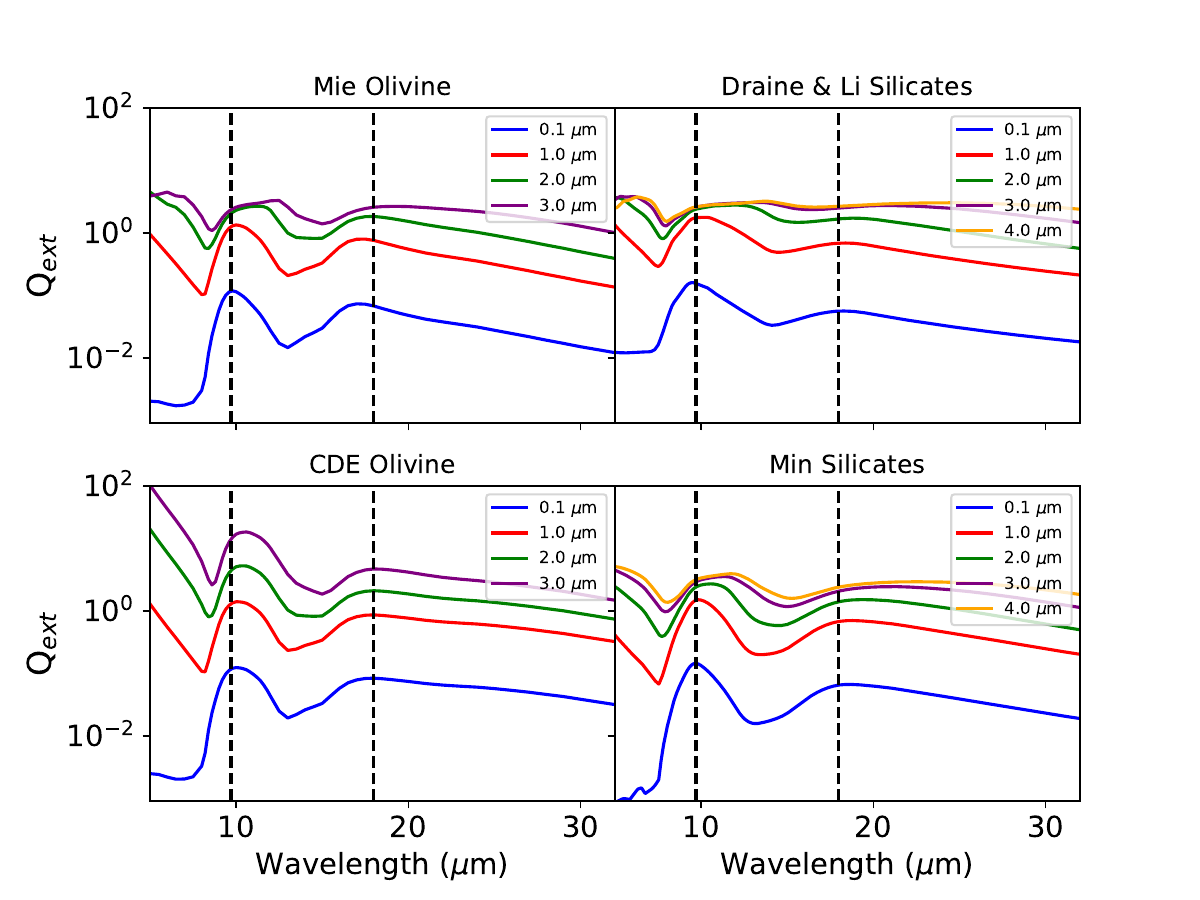}
    \caption{Extinction efficiencies of olivine, DLSil, and MinSil for different grain sizes. The black dashed vertical lines mark the putative peaks of the 10 and $\rm{18\ \mu m}$ silicate features.}
    \label{fig:Qext_silicates_sizes}
\end{figure}

In Figure~\ref{fig:Qext_silicates_sizes}, we show the $Q_{\text{ext}}^{\lambda}$ of CDE and Mie olivine, DLSil, and MinSil for the mentioned grain sizes. It is possible to see that Mie and CDE olivine features look quite similar for 0.1 and  $\rm{1\ \mu m}$ grain sizes. This means that best-fit models featuring (for example) Mie01 should be comparable to fits featuring CDE01 instead. Moreover, the peaks of the $\rm{10}$ and $\rm{18\ \mu m}$ features shift towards longer wavelengths as the grain sizes increase. This effect is more notable in the $\rm{10\ \mu m}$ feature. In the case of DLSil and MinSil, the peaks of both features remain at the same wavelength for grain sizes of 0.1 and $\rm{1\ \mu m}$, while for grain sizes of 2, 3, and $\rm{4\ \mu m}$ the peaks of both features shift towards longer wavelengths as the grain sizes increase. A final highlight of Figure~\ref{fig:Qext_silicates_sizes} is that, for those four dust species, $Q_{\text{ext}}$ flattens as the grain size increases. 

To simplify and shorten the way we refer to the different dust species, we will hereafter use abbreviations made of the name species followed by a number referring to the grain size. Like this, with PoAl, Per, Ol, MgOl, DL, and Min, we refer to porous alumina, periclase, olivine, Mg-rich olivine, Draine \& Li silicates, and Min silicates, respectively. With the numbers 01, 1, 2, 3, and 4 we refer to grain sizes of 0.1, 1, 2, 3, and $\rm{4\ \mu m}$. In the case of olivine, the abbreviation can be preceded by CDE or Mie, referring to the shape of the grain. To summarize, in the subsequent procedure we test the $Q_{\text{ext}}^{\lambda}$ for only one CDE grain size for PoAl, Per, and MgOl; four CDE and four Mie grain sizes for Ol; five Mie grain sizes for DLSil; and five CDE grain sizes for MinSil. The six dust species with their respective grain sizes give us a total of 21 components (excluding those for forsterite and enstaite, see Table~\ref{tab:chemical_compositions}) to be tested in the fits.

\subsection{The fitted combinations} \label{sec:sets_fit_combinations}

In order to explore combinations with a different number of dust species, we test sets made with three to six components. 
Each set includes all the possible combinations given by: 
\begin{equation} \label{eq:combination_formula}
   C_{m,r} = \frac{m!}{(m-r)! r!},
\end{equation}
where $m$ is the total number of species (including combinations of both chemical composition and grain size), and $r$ is the number of components of each set (e.g., for the three-component set, $r=3$).  If we sum the results obtained from this equation for $r=$3,4,5, and 6, we obtain a total of  81,928 performed fits 
for each spectrum of the 49 objects of our sample (see Fig.~\ref{fig:all_spectra}). For each fit, the AIC is calculated (see Appendix~\ref{append:statistical_procedure}). The best-fit model is the one with the minimum value of AIC (AIC$_\text{min}$) among the 81,928 fits performed for each object. However, for the statistical analysis we also take into account those fits with $\Delta$AIC $\lid10$, which we refer to as good fits in Section~\ref{sec:best_fit_combinations} (see Eq.~\ref{eq:delta_AIC}). 


\subsection{Properties derived in the fitting procedure} 
\label{sec:derived_properties}

For all fitted combinations, we calculate the mass fraction of each dust species. In Equation~(\ref{eq:minerology_model}), $c_j$ is the relative number of dust grains or mass fraction of the $j^{\rm th}$ dust species. This value is calculated immediately after a best-fit model is obtained for a given combination of dust species. Technically, given the vector of material densities of the dust species ($\vec{\rho}$) from column~5 in Table~\ref{tab:chemical_compositions}, the best-fit parameter vector ($\vec{P}_{\text{best}}$), and the corresponding covariance matrix ($\bm{P_{\text{cov}}}$), the algorithm returns a vector of mass fractions ($\vec{c}$) and related uncertainties. To calculate this vector, we follow the method presented in Sections 4.3.2, 4.3.3, and 4.3.4 of \citet{Barlow1993}. Then, the mass fraction of the $j$'th dust species is given by
\begin{equation}
    c_j = \frac{\rho_j P_{\text{best},j}}{T},
\end{equation}
where $\rho_j$ and $P_{\text{best},j}$ are the density and the best-fit parameter of the $j$'th dust species, respectively, and $T$ is defined as
\begin{equation}
    T = \vec{\rho} \cdot \vec{P}_{\text{best}}.
\end{equation}

Also, to measure the importance of each dust species in the fits, we calculated the spectral contribution fraction (SCF) of the $j$'th dust species for each fitted combination, as the following:
\begin{equation}
    \text{SCF}_j = \frac{F^{\text{norm}}_{j}}{F^{\text{norm}}_{\text{mod}}}~,
\end{equation}

\noindent where $F^{\text{norm}}_{\text{mod}}$ and $F^{\text{norm}}_{\text{j}}$ are the normalized total fluxes of the fitted model and the $j$'th dust species, respectively. These are defined by:
\begin{equation}
    F^{\text{norm}}_{\text{mod}} = \int  \left( \frac{F_{\lambda,\text{mod}}}{F_{\lambda,\text{c}}} - 1 \right) d\lambda ~,
\end{equation}

\begin{equation}
    F^{\text{norm}}_{j} = \int  \left( \frac{F_{\lambda,j}}{F_{\lambda,\text{c}}} - 1 \right) d\lambda ~.
\end{equation}
where $F_{\lambda,\text{mod}}$ and $F_{\lambda,j}$ are the fluxes of the fitted model and the $j$'th dust species, respectively, and $F_{\lambda,\text{c}}$ is the flux of the fitted continuum (see Section~\ref{sec:continuum}).

\subsection{Caveats} \label{sec:caveats}

The method presented above has a number of caveats and shortcomings due to the inherent approaches applied. Here we summarize them:

\begin{itemize}
    \item The dust species and grain sizes included in this analysis to investigate the AGN dust chemical composition are driven by which optical properties are available in the literature. In this sense, our study may have been limited by the availability of the properties of the chemical compositions. Although out of the scope of this investigations, a broad set of compositions and grain sizes or a subset based on the particularities of the AGN environment, could yield to a different conclusion on the best chemical composition derived here. These limitations need to be addressed by researchers working on the characterization of chemical compositions. 
    
    \item Masking emission lines from the spectra, as they are not included in the modeling, causes the loss of about $\sim30\%$ of data. A possible workaround to this procedure, in an attempt to use as much data as possible, would be to perform a spectral decomposition in which emission lines are simultaneously modeled together with some dust thermal continuum components. However, this decomposition also relies on assumptions on the dust continuum that might bias our results. We have attempted spectrum decomposition with pahfitMCMC \citep{Garcia-Bernete2022} on some sources in our sample. Unfortunately, the silicate feature which is by default included in the models of the code, is unsuited to reproduce our data, resulting in an overfit of PAH feature, in what it is likely an attempt of the model to reproduce the dust continuum. An improvement of the components included in pahfitMCMC might help to mitigate this issue.

    \item Finally, our approach heavily relies on the hypothesis that dust emission in this wavelength range is optically thin. This is supported by the fact that we do observe the silicate features in emission in all of our objects, and that this was one of the main selection criteria for this sample. Nevertheless, this might be only partially true, and might be a source of contamination on our results.
\end{itemize}

While highly desirable, the construction/calculation of a more complete set of dust optical constants to be used in the interpretation of dust emission from the innermost region of AGN, is way beyond the scope of this work. A wider range of grain sizes, chemical compositions better tailored to these particular objects, and physical properties to better represent the harsh environment of AGN, are all characteristics that one would ideally explore trying to better understand AGN dust emission.

\section{Results} \label{sec:results}



\subsection{The good- and best-fit combinations} \label{sec:best_fit_combinations}

\begin{figure}
    \centering
    \includegraphics[scale=1, trim = {0.5cm 0.5cm 0cm 0.5cm}]{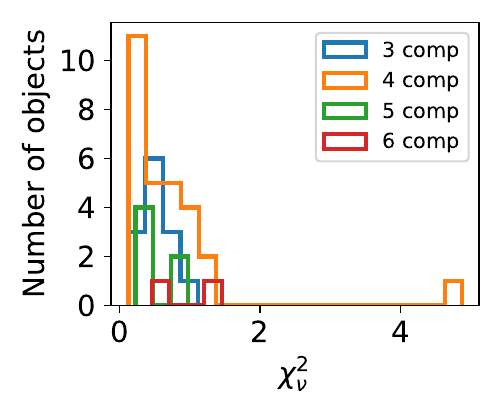}
    \caption{Distribution of the $\chi_{\nu}^2$ values of the best fits with different number of components. Blue, orange, green, and red are the best fits with three, four, five, and six components, respectively.}
    \label{fig:chi2_vs_comp}
\end{figure}

As explained in Section~\ref{sec:sets_fit_combinations}, the best-fit combination selected for each spectrum is the one with the minimum AIC. In Table~\ref{tab:best_fits_sample} we present the best-fit combination for each object, the mass and spectral contribution fractions per component for each case, and the $\chi^2_{\nu}$ of each fit. 
We can see that PoAl01 has the largest mass fraction in most cases, with a median value of 0.55, followed by Per01 with a median of 0.22. PoAl01 and Per01 are also the species with the largest SCF, with a median of 0.39 and 0.30, respectively. The number of components (or dust species) required for each model to fit the MIR spectra varies depending on the object. To illustrate the latter result, Fig.~\ref{fig:chi2_vs_comp} shows the distribution of the $\chi^2_{\nu}$ values of the best fits, for models using a different number of components. The best fits for the spectra of all objects except one (Mrk~1210), have $\chi^2_{\nu} <2$ with a median of 0.50. It is notable that the majority (57\%) of the best fits are combinations composed of four dust species. Models adopting three, five, and six components are required only for 27\%, 12\%, and 4\% of the best fits, respectively.

\begin{table*}
\begin{threeparttable}
\small 
\renewcommand{\arraystretch}{0.9}
\begin{center}
\caption{\normalsize Best-fit combination of our studied sample and derived properties.}
\label{tab:best_fits_sample}
\begin{tabular}{lcccc} \hline \hline
Object name & Best-fit combination & Mass fractions & Spectral contribution fractions & $\chi^2_{\nu}$  \\ 
(1) & (2) & (3) & (4) & (5) \\ \hline
2MASX~J11454045-1827149 & PoAl01/Per01/MieOl01/MieOl3 & 0.55/0.17/0.28/0.004 & 0.49/0.32/0.14/0.05 & 0.13 \\
3C~120 & PoAl01/Per01/MieOl2/CDEOl01 & 0.51/0.20/0.02/0.27 & 0.35/0.28/0.23/0.15 & 0.26 \\
Ark~120 & PoAl01/Per01/MieOl3/CDEOl1& 0.58/0.28/0.02/0.13/ & 0.23/0.23/0.09/0.45  & 0.29 \\
Ark~347 & PoAl01/Per01/CDEOl1/DL2  & 0.48/0.43/0.07/0.02 & 0.21/0.38/0.25/0.26 & 0.57\\
CGCG~420-015 & PoAl01/Per01/CDEOl1/DL2 & 0.60/0.36/0.03/0.01 & 0.33/0.41/0.15/0.12 & 1.11 \\
ESO~138-G001 & PoAl01/Per01/MieOl01 & 0.48/0.11/0.40 & 0.51/0.25/0.25 & 0.46 \\
ESO~141-G055 & PoAl01/Per01/MieOl2/MieOl3/CDEOl1 & 0.48/0.36/0.02/0.03/0.11 & 0.17/0.25/0.10/0.14/0.35 & 0.34  \\
ESO~374-G044 & PoAl01/Per01/MieOl01 & 0.75/0.14/0.11 & 0.68/0.26/0.06 & 1.05 \\
ESO~511-G030 & PoAl01/Per01/MieOl2/CDEOl1 & 0.71/0.22/0.01/0.05 & 0.39/0.25/0.09/0.27 & 0.29 \\
ESO~548-G081 & PoAl01/Per01/MieOl01/MieOl3/CDEOl01 & 0.33/0.16/0.24/0.01/0.27 & 0.29/0.29/0.12/0.10/0.20 & 0.42 \\
Fairall~51 & PoAl01/Per01/MieOl2/CDEOl01 & 0.38/0.13/0.05/0.44 & 0.24/0.17/0.37/0.23 & 0.93 \\
FAIRALL~9 & PoAl01/Per01/CDEOl1/DL2 & 0.55/0.37/0.06/0.01 & 0.26/0.36/0.27/0.11 & 0.60 \\
IC~4329A & PoAl01/Per01/MgOl01/MieOl3 & 0.65/0.20/0.14/0.01 & 0.50/0.31/0.09/0.10 & 0.50 \\
II~SZ~010 & PoAl01/Per01/MieOl3/DL01 & 0.57/0.22/0.002/0.21 & 0.49/0.38/0.03/0.10 & 1.12 \\
II~Zw~136 & PoAl01/Per01/CDEOl3 & 0.71/0.29/0.002 & 0.51/0.42/0.07 & 0.19 \\
I~Zw~1 & PoAl01/CDEOl01/DL01 & 0.26/0.39/0.35 & 0.33/0.42/0.25 & 0.12 \\
M~106 & PoAl01/MieOl2/DL01 & 0.73/0.01/0.26 & 0.68/0.18/0.14 & 0.73 \\
MCG~-01-13-025 & PoAl01/Per01/MieOl2/CDEOl3 & 0.58/0.37/0.04/0.004 & 0.28/0.37/0.25/0.11 & 0.19 \\
MCG~+04-22-042 & PoAl01/Per01/MgOl01/MieOl3 & 0.48/0.16/0.35/0.003 & 0.42/0.29/0.25/0.04 & 0.32 \\
MCG~-06-30-015 & Per01/MieOl3/CDEOl01 & 0.76/0.03/0.21 & 0.72/0.20/0.08 & 0.40 \\
Mrk~1018 & PoAl01/Per01/MieOl2/CDEOl01 & 0.45/0.17/0.006/0.38 & 0.38/0.29/0.07/0.27 & 0.16 \\
Mrk~110 & PoAl01/Per01/MieOl01/MieOl2 & 0.46/0.23/0.31/0.004 & 0.40/0.40/0.16/0.05 & 1.04 \\
Mrk~1210 & PoAl01/Per01/MieOl3/CDEOl01 & 0.62/0.27/0.01/0.10 & 0.44/0.39/0.11/0.06 & 4.88 \\
Mrk~1392 & PoAl01/Per01/MieOl01/MieOl3 & 0.59/0.12/0.29/0.004 & 0.56/0.23/0.16/0.06 & 0.53 \\
Mrk~1393 & PoAl01/Per01/MieOl1 & 0.74/0.24/0.02 & 0.56/0.37/0.08 & 0.53 \\
Mrk~279 & PoAl01/Per01/MieOl01 & 0.45/0.28/0.27 & 0.39/0.48/0.14 & 0.44 \\
Mrk~290 & PoAl01/Per01/MieOl3/CDEOl01 & 0.51/0.23/0.007/0.25 & 0.39/0.37/0.08/0.17 & 0.64 \\
Mrk~417 & PoAl01/Per01/MieOl01 & 0.70/0.25/0.06 & 0.56/0.41/0.03 & 0.62 \\
Mrk~590 & PoAl01/Per01/CDEOl01/DL4 & 0.61/0.10/0.28/0.005 & 0.53/0.18/0.20/0.09 & 0.24 \\
Mrk~705 & PoAl01/Per01/MieOl01 & 0.71/0.11/0.18/ & 0.68/0.22/0.10 & 0.76 \\
Mrk~841 & PoAl01/Per01/MieOl01/MieOl3 & 0.69/0.19/0.12/0.003 & 0.58/0.33/0.06/0.03 & 0.48 \\
NGC~1052 & PoAl01/Per01/MieOl3/CDEOl01 & 0.60/0.18/0.01/0.21 & 0.47/0.28/0.11/0.14 & 0.68 \\
NGC~1275 & PoAl01/Per01/MgOl01/MieOl3/CDEOl01 & 0.55/0.04/0.20/0.002/0.21 & 0.56/0.08/0.17/0.02/0.18 & 0.43 \\
NGC~3783 & PoAl01/Per01/MieOl3/CDEOl01/DL4 & 0.32/0.44/0.03/0.19/0.03 & 0.14/0.38/0.15/0.07/0.26 & 0.92 \\
NGC~4151 & PoAl01/Per01/MgOl01/MieOl01/MieOl3 & 0.62/0.18/0.12/0.07/0.003 & 0.54/0.31/0.09/0.04/0.03 & 0.87 \\
NGC~4507 & PoAl01/Per01/MieOl01/MieOl2 & 0.50/0.32/0.16/0.18 & 0.34/0.44/0.06/0.16 & 0.79 \\
NGC~4939 & PoAl01/Per01/MieOl01 & 0.56/0.21/0.23 & 0.50/0.38/0.12 & 0.61 \\
NGC~526A & PoAl01/Per01/DL01/Min4 & 0.66/0.25/0.09/0.007 & 0.39/0.31/0.03/0.27 & 0.73 \\
NGC~5548 & PoAl01/Per01/CDEOl1/DL2 & 0.64/0.30/0.05/0.01 & 0.33/0.31/0.25/0.11 & 1.25 \\
NGC~6814 & PoAl01/Per01/MieOl01/MieOl2 & 0.51/0.08/0.37/0.04 & 0.37/0.12/0.16/0.35 & 0.28 \\
NGC~7213 & PoAl01/Per01/MieOl2/CDEOl1/CDEOl3/DL2 & 0.46/0.41/0.02/0.08/0.008/0.03 & 0.15/0.27/0.07/0.22/0.13/0.16 & 1.32 \\
PG~0804+761 & PoAl01/Per01/CDEOl1/DL01/Min3 & 0.33/0.23/0.15/0.26/0.03 & 0.09/0.13/0.36/0.04/0.37 & 0.23 \\
PG~1211+143 & PoAl01/Per01/MieOl3/CDEOl1 & 0.44/0.39/0.008/0.17 & 0.15/0.28/0.04/0.53 & 1.23 \\
PG~1351+640 & PoAl01/Per01/MieOl01/CDEOl1/DL01/Min3 & 0.12/0.11/0.37/0.08/0.31/0.008 & 0.06/0.12/0.11/0.39/0.09/0.22 & 0.46 \\
PG~1448+273 & PoAl01/Per01/MieOl3/DL01 & 0.62/0.22/0.005/0.15 & 0.50/0.37/0.06/0.07 & 0.66 \\
PG~2304+042 & PoAl01/Per01/MieOl1 & 0.72/0.23/0.05 & 0.49/0.33/0.19 & 0.43 \\
PICTOR~A & PoAl01/Per01/MieOl3/CDEOl01 & 0.50/0.14/0.003/0.35 & 0.45/0.25/0.04/0.26 & 0.28 \\
UGC~6728 & PoAl01/Per01/MieOl01/MieOl3 & 0.37/0.20/0.42/0.007 &  0.33/0.37/0.22/0.08 & 0.18 \\
UM~614 & PoAl01/Per01/CDEOl1 & 0.70/0.27/0.03 & 0.45/0.36/0.19 & 0.13 \\
\hline\hline
\end{tabular}
\begin{tablenotes}
\item \noindent \textbf{Notes:} Column 1 is the name of the object. Column 2 is the combination of dust species that best fit the observed MIR spectrum. Columns 3 and 4 show the mass and spectral contribution fractions of the dust species from the best fit. Column 5 is the $\chi^2_{\nu}$ obtained with the best fit. 
\end{tablenotes}
\end{center}
\end{threeparttable}
\end{table*}

\begin{figure*}
    \centering
    \includegraphics[scale=0.32, trim = {1.5cm 0cm 0cm 0cm}]{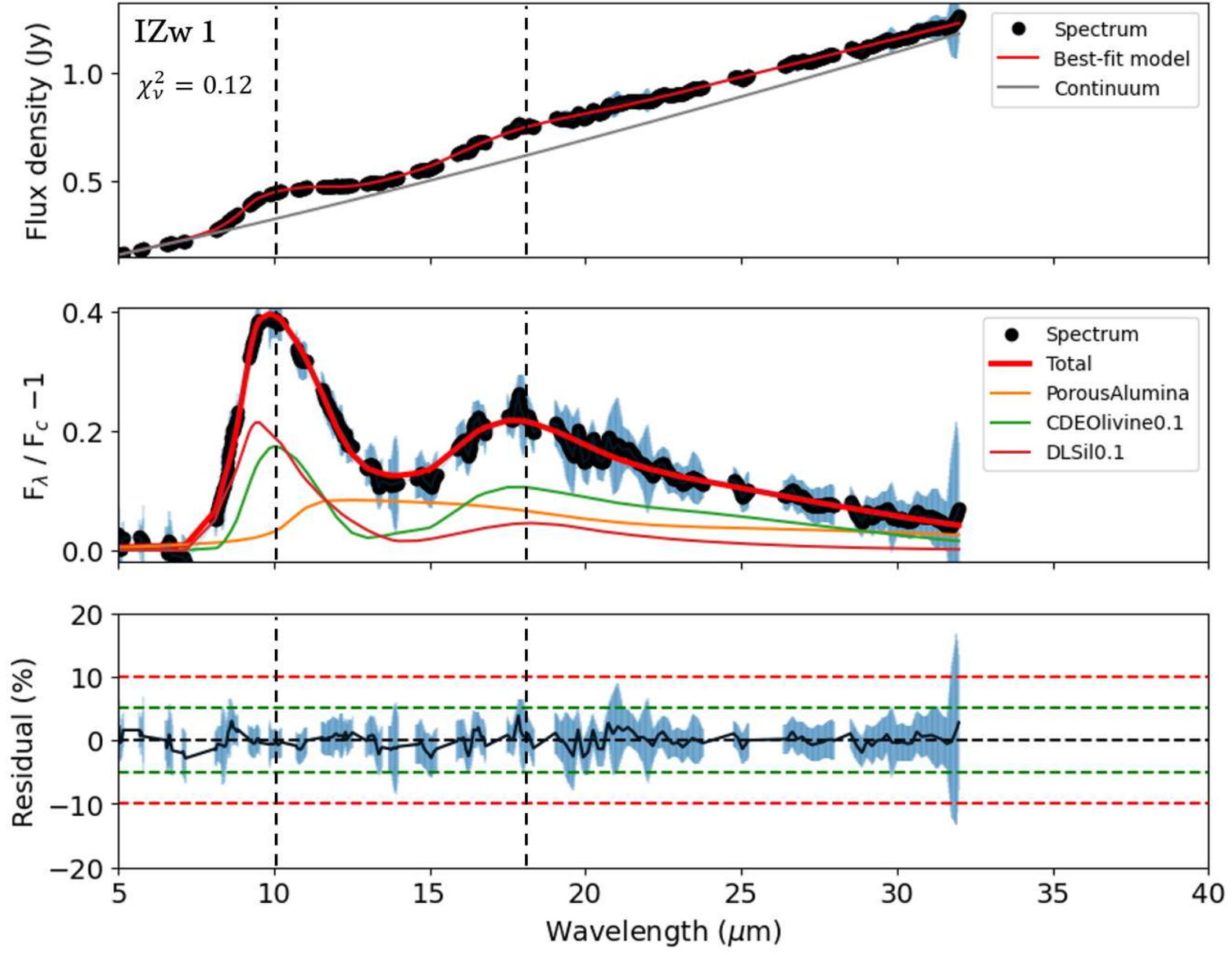}
    \includegraphics[scale=0.32, trim = {1.5cm 0cm 1.5cm 0cm}]{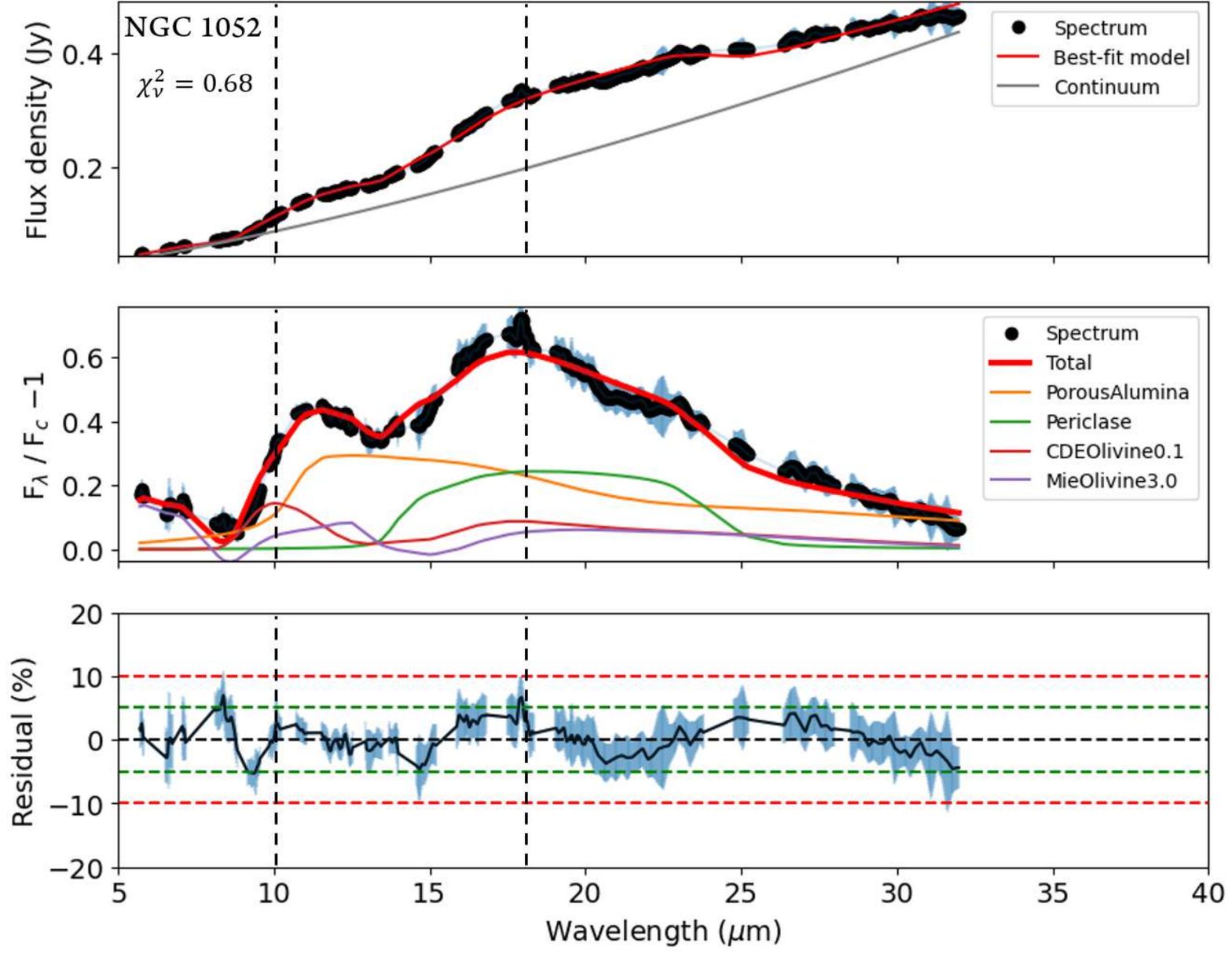}
    \caption{Left: Best fit of the MIR spectrum of I~Zw~1 with a combination composed of PoAl01, Per01, CDEOl01, and DL01. Right: Best fit of the MIR spectrum of NGC~1052 with a combination composed of PoAl01, Per01, CDEOl01, and MieOl3. The upper panel shows the data (black points), their error bars (in blue), the best fit (red solid line), and the continuum (gray solid line) used in that best fit. The middle panel shows the data (black points), their errors (in blue), the fitted dust species (in colors), and the best fit (red thick solid line). The y-axis in this panel is the total flux normalized by the continuum flux minus one in order to highlight the silicate features and the spectral contributions of the dust species. The lower panel shows the residual normalized by the best-fit model in percentage (black solid line) and thresholds at 5\% and 10\% (green and red dashed lines, respectively). The black dashed vertical lines are at 10 and $\rm{18\ \mu m}$.}
    \label{fig:best_fits_examples}
\end{figure*}

\begin{figure}
    \centering
    \includegraphics[scale=0.32, trim = {1.5cm 0cm 0cm 0cm}]{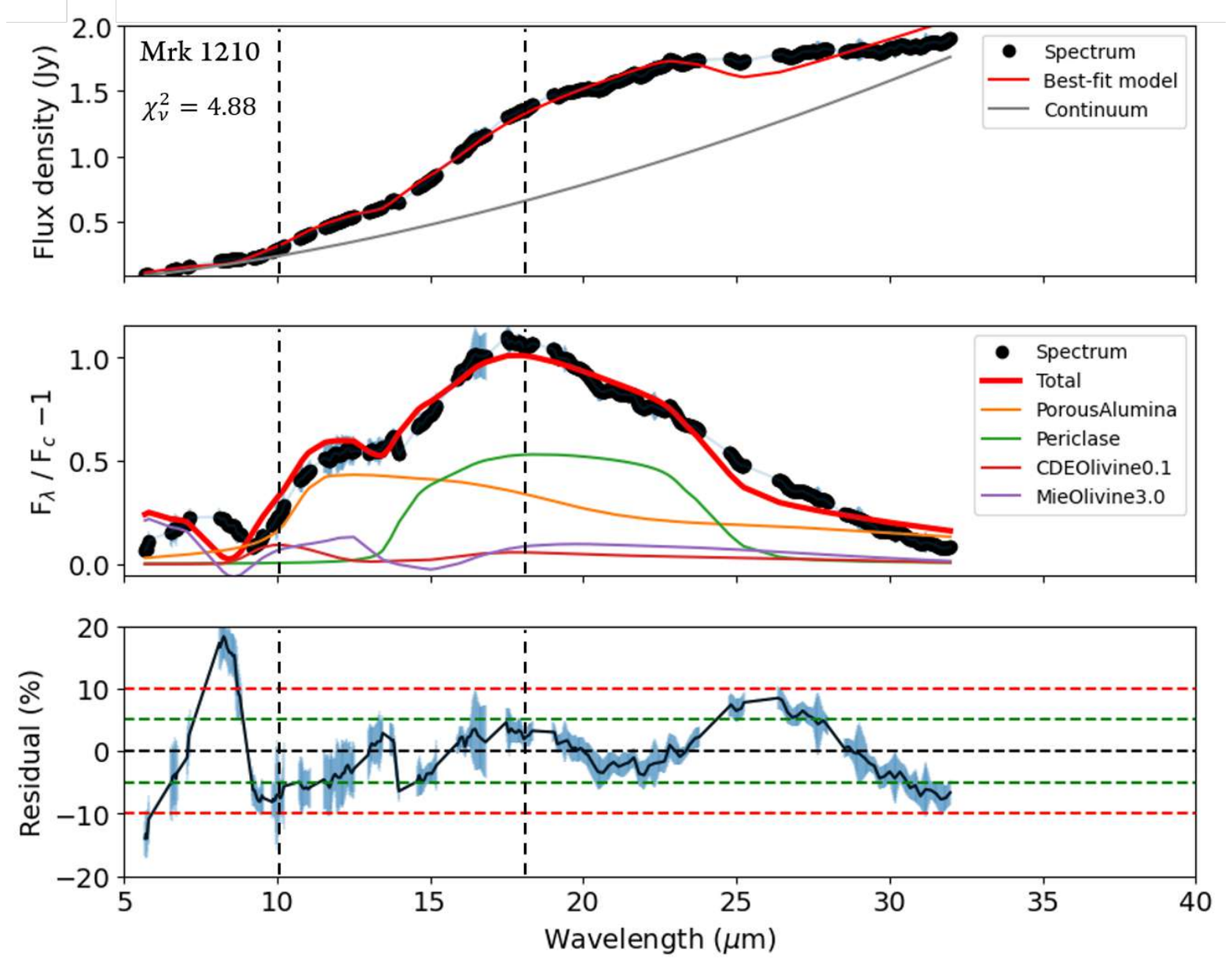}
    \caption{Left: Best fit of the MIR spectrum of Mrk~1210 with a combination composed of PoAl01, Per01, CDEOl01, and MieOl3. The description is same as Fig.~\ref{fig:best_fits_examples}}
    \label{fig:best_fit_Mrk1210}
\end{figure}

As examples of the best fits obtained, Figure~\ref{fig:best_fits_examples} shows the best-fit models for the spectra of I~Zw~1 and NGC~1052. We show the fits for these objects because I~Zw~1 has the best fit with the lowest $\chi^2_{\nu}$ value and NGC~1052 is a good example to show the behavior of some dust species, which is explained later. The residuals between the observed spectra and the model for both objects are within $5\%$ at all wavelengths and their $\chi^2_{\nu}$ are 0.12 and 0.68, respectively. In the case of I~Zw~1, its low $\chi^2_{\nu}$ value might indicate that the model has too many free parameters or that the errors are overestimated. The number of free parameters of this particular best-fit model is only $k=7$ (four for the poly3 continuum and three for the dust species), which is the minimum number of free parameters used in this analysis. Therefore, it seems that an overestimation of the observed flux uncertainties is more plausible for this particular case and others with low $\chi^2_{\nu}$. Indeed, this is the case for many \emph{Spitzer}/IRS spectra analysed in other works as well \citep[e.g.][]{Gonzalez-Martin2019II,Gonzalez-Martin2023}. In the case of NGC~1052, we can see that CDEOl01 contributes to fit the peak of the $\rm{18\ \mu m}$ silicate feature shifted towards shorter wavelengths and MieOl3 contributes to fit the peak of the $\rm{10\ \mu m}$ silicate feature shifted towards longer wavelengths. In addition, we show in Figure~\ref{fig:best_fit_Mrk1210} the best fit of the spectrum of Mrk~1210 because it is the object with the largest $\chi^2_{\nu}$ value among those of the sample we analysed, and the only one with $\chi^2_{\nu}>2$. It shows residuals within $5-10\%$ at $\rm{\lambda>10\ \mu m}$ but at $\rm{\lambda<10\ \mu m}$ the model does not fit the data well, which is the reason for its $\chi^2_{\nu}$ value of 4.88.

\begin{figure}
    \centering
    \includegraphics[scale=0.4, trim = {1.6cm 0.5cm 0cm 0cm}]{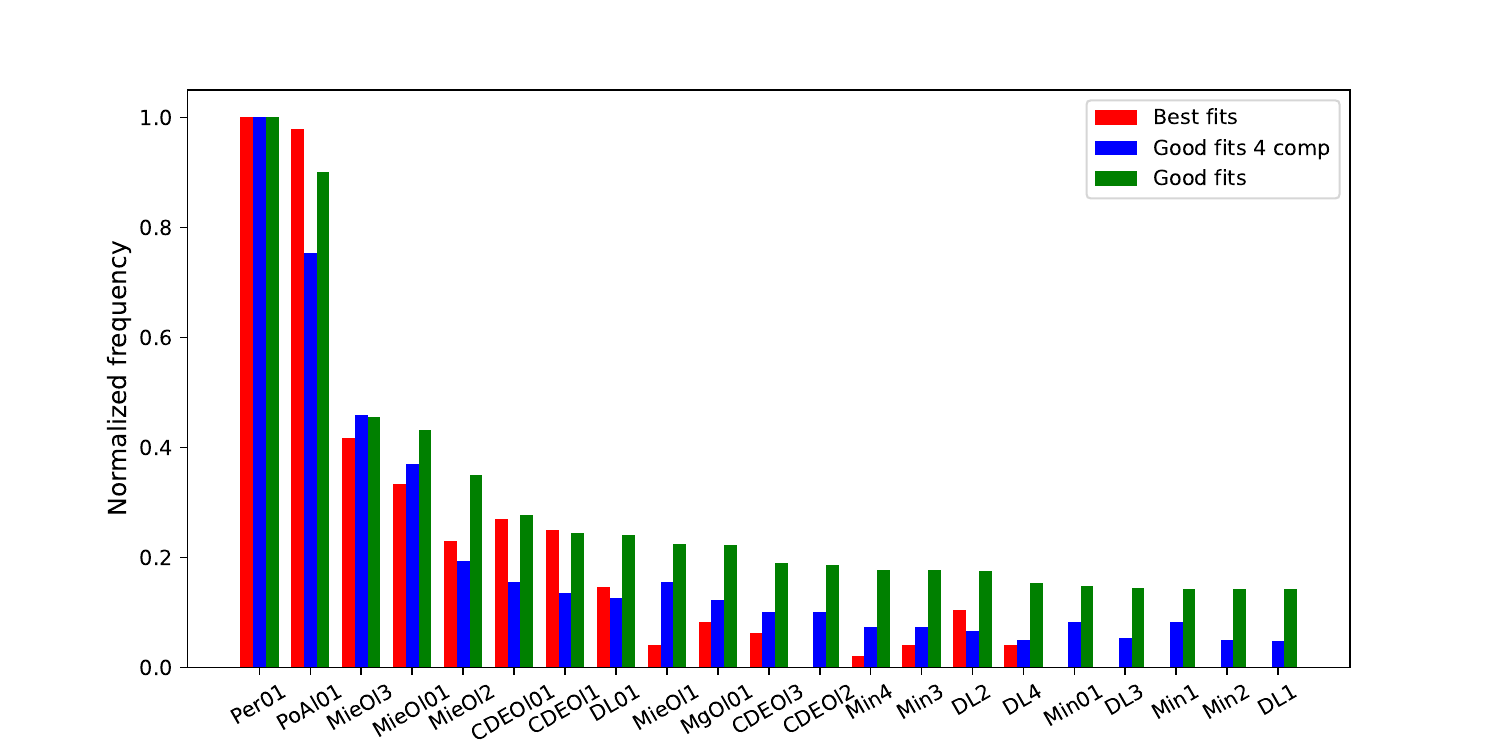}
    \caption{Normalized frequency of the dust species used in the best fits (red), good fits with four components (blue), and good fits (green) of the 49 AGN. The normalization values are 48, 316, and 17825, respectively.}
    \label{fig:bar_plot_dust_species}
\end{figure}

Taking into account only the best fits from the tens of thousands of fits performed for each object, we find that, from the 21 possible dust species considered for fitting the spectra (see Section~\ref{sec:sets_fit_combinations}), 15 are used (CDEOl3, Min01, DL3, Min1, Min2, and DL1 do not feature in any of the best-fit combinations). This result implies that a larger set of grain size/shape distributions need to be considered to reproduce the variety of spectra seen in AGN. This is in addition to requiring two different ``types'' of dust — oxides and amorphous silicates. The latter is required to reproduce the 10 and 18 µm features, while the former are needed for the broader features. The red bars of Figure~\ref{fig:bar_plot_dust_species} show the normalized frequency of the dust species used in the best fits. We can see that PoAl01 and Per01 are needed to obtain the best fit for almost all objects. These two species contribute to reproduce the silicate feature at $\rm{10\ \mu m}$ (with a minor contribution to the one at $\rm{18\ \mu m}$, see I~Zw~1 in Fig.~\ref{fig:best_fits_examples}), and the broad feature in the range between 15 and $\rm{25\ \mu m}$ (see Mrk~1210 in Fig.~\ref{fig:best_fit_Mrk1210}), respectively. PoAl01 is not needed to obtain the best fit for the spectrum of MCG~-06-30-015 because its $\rm{10\ \mu m}$ feature strength is much lower with respect to its $\rm{18\ \mu m}$ counterpart, while Per01 is not needed in the spectra of I~Zw~1 and M~106 because they do not show the broad feature between 15 and $\rm{25\ \mu m}$. MieOl01 and MieOl3 are the dust species most frequently appearing in the best fits after the two mentioned above. These dust species, in particular MieOl3, are relevant to explain the shift towards longer wavelengths of the $\rm{10\ \mu m}$ silicate emission feature and to contribute to the NIR excess below $\rm{7\ \mu m}$ (see Fig.~\ref{fig:best_fits_examples}).

\begin{figure}
    \centering
    \includegraphics[scale=0.6, trim = {0.5cm 0.5cm 0cm 0cm}]{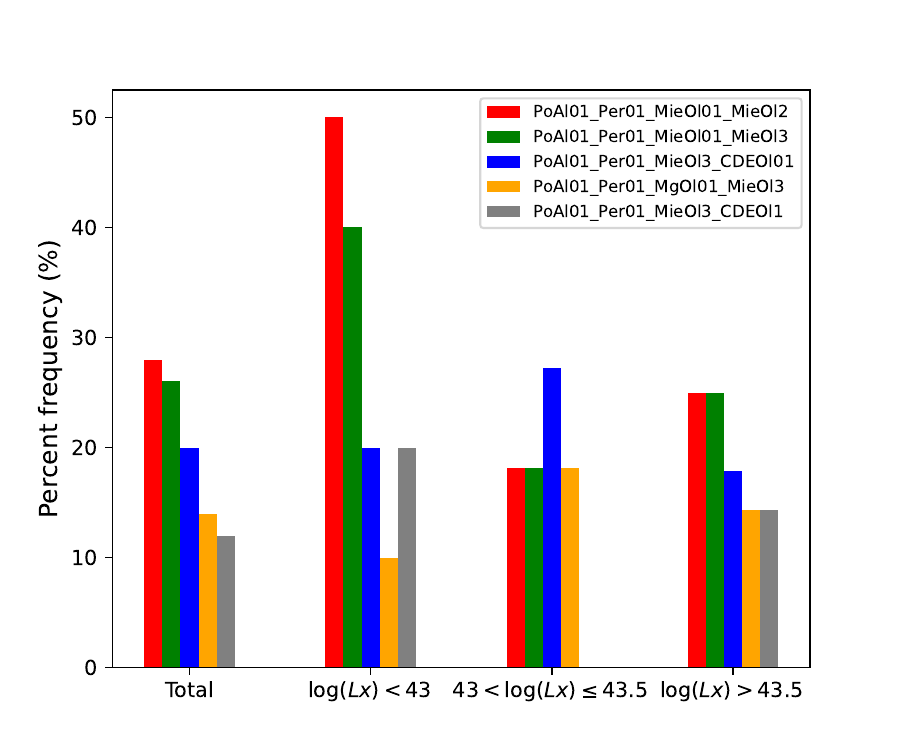}
    \caption{ Percent frequency of the five most repeated good-fit combinations with four components among the 49 AGN. The bars are grouped in four sets with different number of objects. From left to right, the total number of objects, objects with $\rm{log(L_{X})<43}$, $\rm{43<log(L_{X})\leq 43.5}$, and  $\rm{log(L_{X})>43.5}$}, respectively.
    \label{fig:Bar_plot_good_fits}
\end{figure}

We obtained a total of 24,740 good-fit combinations (i.e. with $|\Delta$AIC$|<10$) for the whole sample. The blue and green bars of Figure~\ref{fig:bar_plot_dust_species} show the normalized frequency of the dust species used in the good-fit combinations with four components and the total good fits, respectively. We can note that PoAl01, Per01, MieOl3, and MieOl01 are also the most used dust species in the good fits,  mimicking the results obtained for the best fits. In order to look for a combination of dust species which is more likely to provide a good spectral fit, we grouped them according to the number of components and we analysed the most frequent good-fit combinations of the group with four components since this is the number of components having the majority of the best fits. We found that the five most frequent good-fit combinations have a higher occurrence than the rest. These are shown in Figure~\ref{fig:Bar_plot_good_fits}, where we can see that they are composed, in general, of $\rm{0.1\ \mu m}$ porous alumina (PoAl01), $\rm{0.1\ \mu m}$ periclase (Per01), and either Mie or CDE olivine (MieOl or CDEOl) with different grain sizes. These five combinations contain one small-grain olivine component and one large-grain olivine component. This is necessary to fit the shape of the 10 and $\rm{18\ \mu m}$ features. This figure also shows the percentage of objects with $\rm log(L_{X})$ within the ranges $\rm{log(L_{X})>43.5}$, $\rm{43<log(L_{X})\leq 43.5}$, and  $\rm{log(L_{X})<43}$ that have these five combinations as good fits. We can see that none of the X-ray luminosity ranges prefer a particular good-fit combination since the five ones are good fits for almost the same percentage of objects in the three ranges. This is a result of the fact that the fits cannot significantly differentiate between grain sizes that are relatively close to each other. As we mentioned before, the five combinations are a ``small + large'' combination, e.g., the green combination has MieOl01 and MieOl3, while the blue combination has CDEOl01 and MieOl3. Since Figure~\ref{fig:Qext_silicates_sizes} shows that CDEOl01 is very similar to MieOl01, there is no statistically significant difference in the populations represented by these two different colors.

\subsection{The best chemical composition} \label{sec:best_chemical_composition}

Despite the fact that the chemical composition (i.e., the relative fractions of the individual components of the dust), the dust species, and the grain size/shape seem to differ from object to object (see Table~\ref{tab:best_fits_sample}), we have found that the dust species that are most appropriate to represent the chemical composition of our AGN sample is the combination of PoAl01, Per01, MieOl01, and MieOl3. To get to this particular combination, we have taken into account the statistical frequency of each dust species in the best fits (see Fig.~\ref{fig:bar_plot_dust_species}) and the fact that more than 50\% of the best fits are combinations of four dust species (see Fig.~\ref{fig:chi2_vs_comp}). The aforementioned composition turns out to be the best  compromise to reproduce the silicate feature emission in our dataset, and is hence the one we propose to be used when calculating radiative transfer models of the dusty torus emission. A more complete analysis would require to produce a grid of RT models with a complete and homogeneous covering of both the ``standard'' dust parameters, but also exploring the various combinations of dust properties. This approach, though, is far from optimal as it would exponentially increase the number of models.



\begin{figure}
    \centering
    \includegraphics[scale=1, trim = {0.5cm 0.5cm 0cm 0.5cm}]{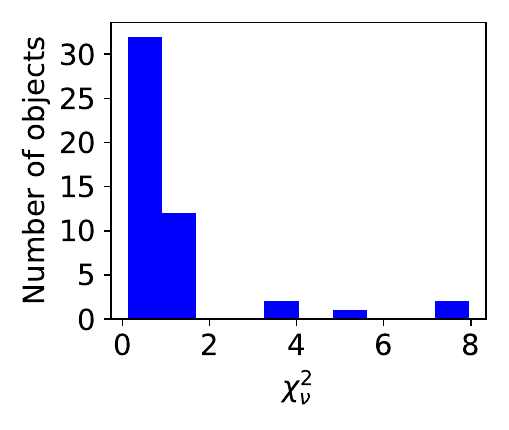}
    \caption{Distribution of the $\chi_{\nu}^2$ values of fits with the particular combination composed of PoAl01, Per01, MieOl01, and MieOl3.}
    \label{fig:chi2_final_comp}
\end{figure}

\begin{figure*}
    \centering
    \includegraphics[scale=0.5, trim = {0cm 0cm 0cm 0cm}]{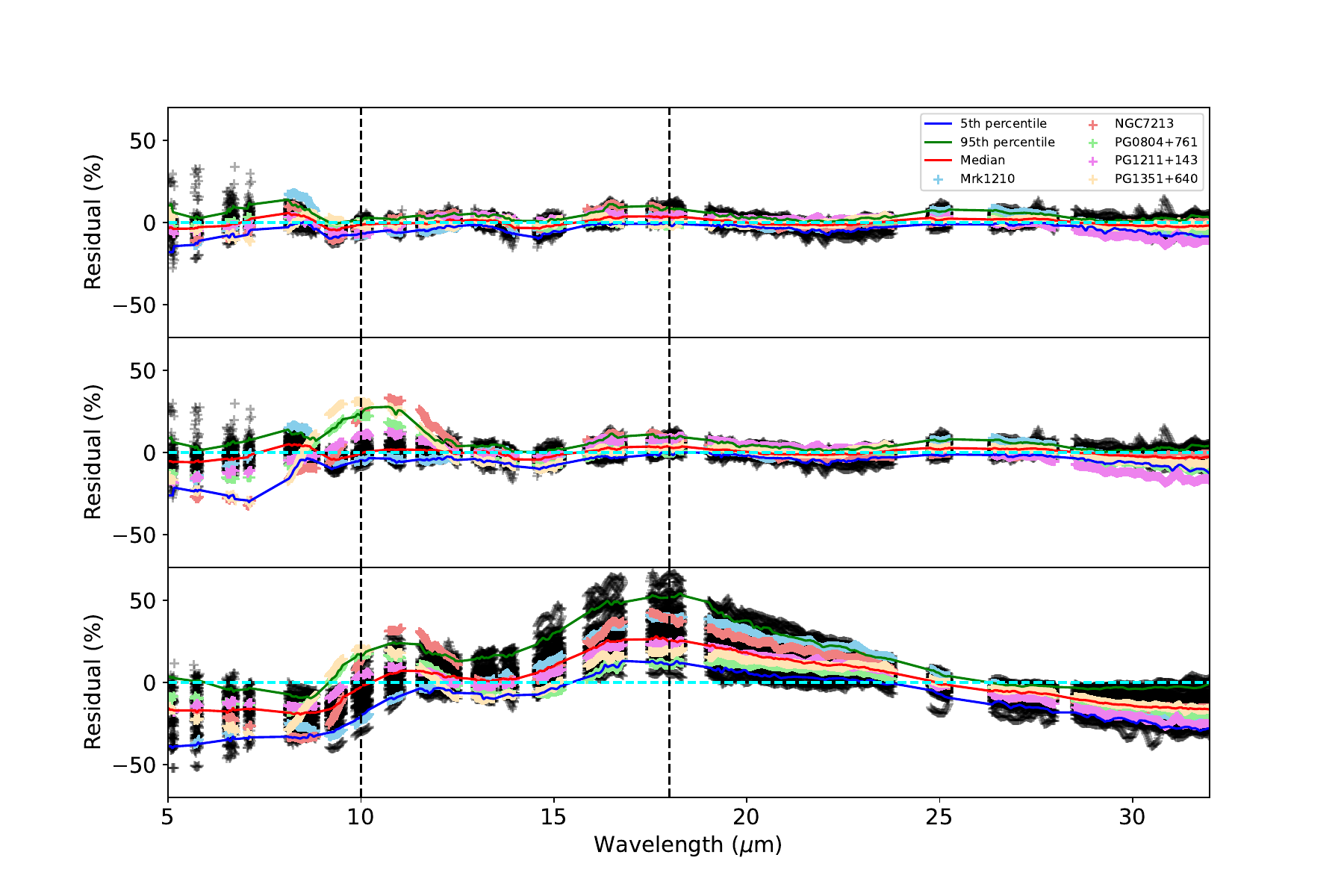}
    \caption{Residuals of the spectral fits normalized by the fit model for our 49 objects. The top panel shows the residuals of the best fits. The middle panel shows the residuals of the combination composed of PoAl01, Per01, MieOl01, and MieOl3. The bottom panel shows the residuals of the fits using the astronomical silicates DLSil. The red solid line shows the median value. The green and blue solid lines represent the 5th and 95th percentiles of the distribution. The cyan dashed line indicates the zero value residual. The colored crosses are the five objects that are outliers in Figure~\ref{fig:chi2_final_comp} because they show} $\chi^2_{\nu}>2$. .
    \label{fig:all_residuals}
\end{figure*}

We have analysed how well the particular combination composed of PoAl01, Per01, MieOl01, and MieOl3 is, by fitting the spectra of our sample. Figure~\ref{fig:chi2_final_comp} shows the distribution of the $\chi^2_{\nu}$ values of fits with this specific combination (having a median $\chi^2_{\nu}=0.65$). We can see that  five objects (Mrk~1210, NGC~7213, PG~0804+761, PG~1211+143, and PG~1351+640) are outliers in this distribution with $\chi^2_{\nu}>2$. The middle panel of Figure~\ref{fig:all_residuals} shows the residuals of the spectral fits for our 49 objects using this combination, and it highlights these five objects. Comparing these residuals with the ones obtained in the best fits (top panel), we can see that the main difference is that the latter have larger ones ($>10\%$) at $9~\mu{\text{m}} < \lambda < 12~\mu{\text{m}}$ for NGC~7213, PG~0804+761, PG~1211+143 and PG~1351+640. For Mrk~1210, the residuals of the particular combination are not very different from the ones of the best fit. This is because both compositions are very similar, the only difference is that the best fit includes CDEOl01 instead of MieOl01, which have almost the same behavior (see Fig.~\ref{fig:Qext_silicates_sizes}). For NGC~7213, PG~0804+761, PG~1211+143, and PG~1351+640, the residuals of the particular combination differ from the ones of the best fit mainly at $8~\mu{\text{m}} < \lambda < 13~\mu{\text{m}}$, meaning that the former can not reproduce well the $\rm{10\ \mu m}$ silicate feature in the spectra of these objects. To obtain an acceptable fit model, NGC~7213 needs MieOl2, CDEOl1, CDEOl3, and DL2 instead of MieOl01 and MieOl3; PG~0804+761 needs CDEOl1, DL01, and Min3 instead of MieOl01 and MieOl3; PG~1211+143 needs CDEOl1 instead of MieOl01; and PG~1351+640 needs CDEOl1, DL01, and Min3 instead of MieOl3.

\begin{figure}
    \centering
    \includegraphics[scale=0.4, trim = {1cm 0cm 0cm 0cm}]{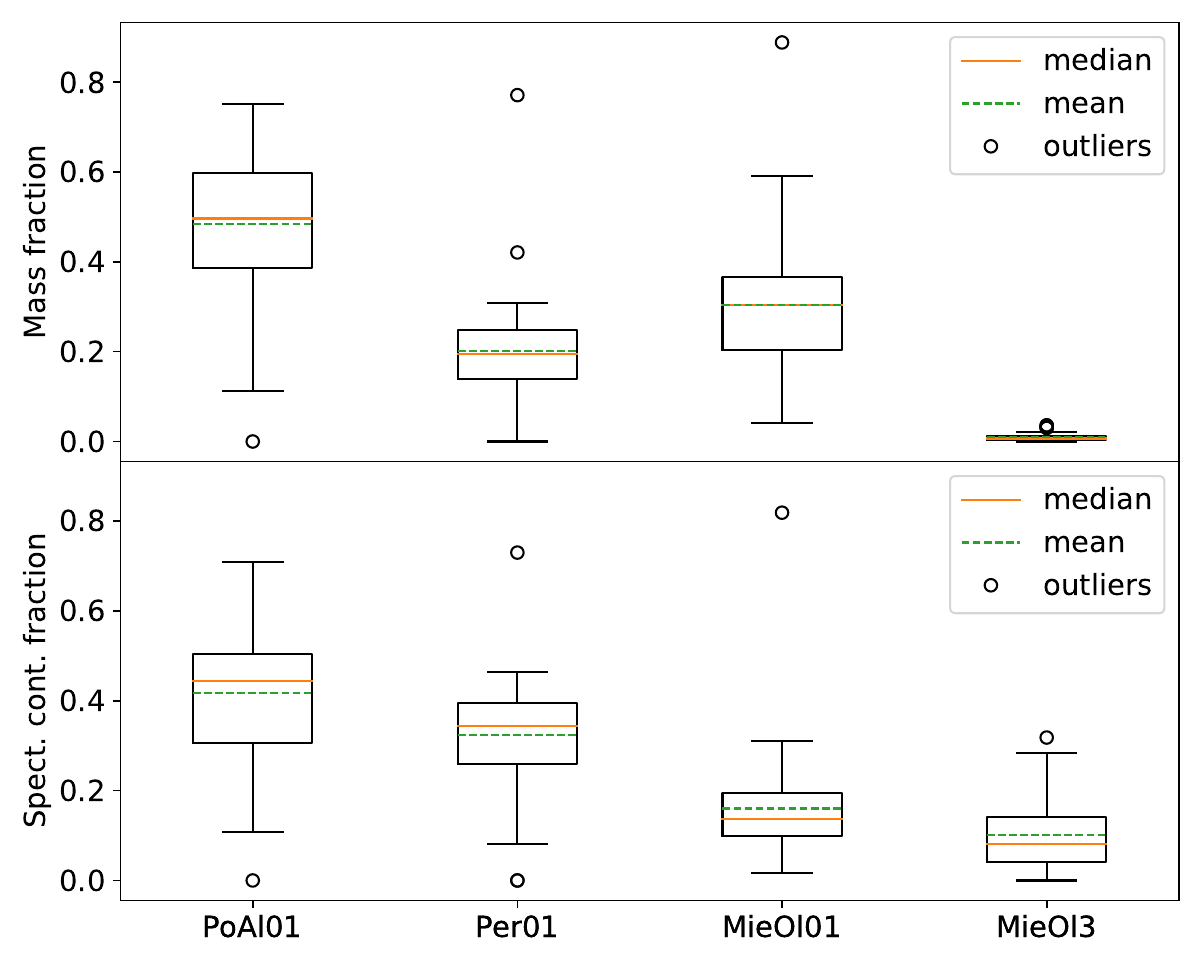}
    \caption{Top panel shows the distributions of the mass fractions of the four dust species that compose our proposed chemical composition. Bottom panel shows the distributions of their spectral contribution fractions.
    Each box extends from the first to the third quartile (inter-quartile range, IQR) of the data. Green dashed and orange solid lines are the mean and the median, respectively. The whiskers extend from the box by 1.5 times the IQR (i,e., 1.5 times the box size). Empty circles are outliers, i.e., cases where the value is out of the extension of the whiskers.} 
    \label{fig:box_plot_final_composition}
\end{figure}

Figure~\ref{fig:box_plot_final_composition} shows box plots of the distributions of the mass and spectral contribution fractions of each of the four dust species of the particular combination. 
In terms of mass, PoAl01 and MieOl01 are dominant in the dust composition having the largest median values of $\sim0.5$ and $\sim0.3$, respectively, while Per01 and MieOl3 have median values of $\sim0.2$ and $\sim0.007$, respectively. On the other hand, in terms of spectral flux, PoAl01 and Per01 are the most important with a median spectral contribution fraction of around 0.40, while MieOl01 and MieOl3 have a median spectral contribution fraction of around 0.10. Despite its very low mass fraction, MieOl3 contributes significantly to the shape of the spectrum (see Sec.~\ref{sec:the_anomalous}). 

\begin{figure}
\centering
\includegraphics[scale=0.6, trim = 0cm 0cm 0cm 0cm]{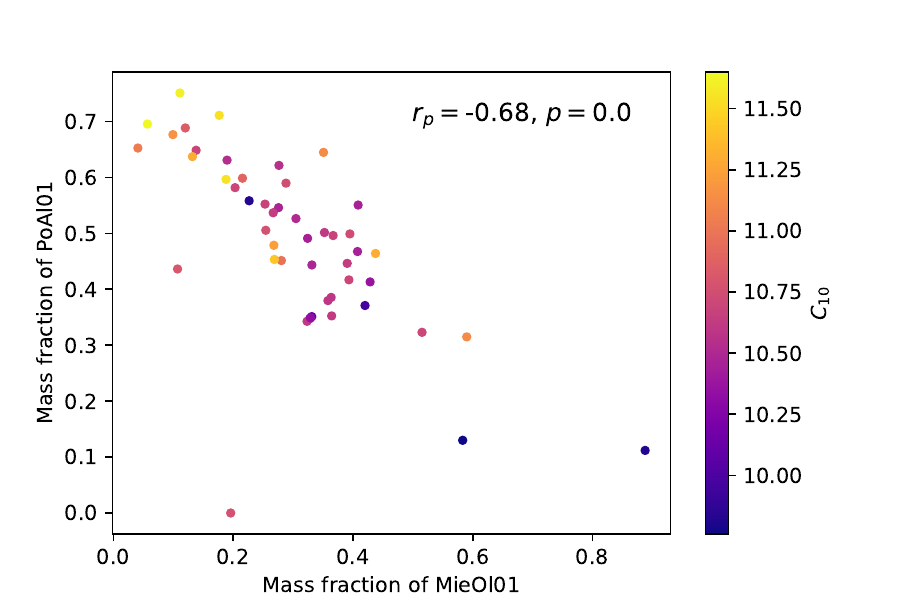}
\caption{Mie olivine of $0.1\mu$m mass fractions versus porous alumina $0.1\mu$m mass fraction. The color bar represents the $10\mu$m silicate feature peak wavelength.}
\label{fig:wave_vs_massfrac_compress}
\end{figure}

In addition, through the Pearson coefficient ($r_p$), we have explored correlation between parameters such as mass fraction, spectral contribution fraction, the $\rm{10\ \mu m}$ silicate feature strength ($\rm{\rm{S_{10\mu m}}}$), the $\rm{10\ \mu m}$ feature wavelength peak ($\rm{\rm{C_{10\mu m}}}$), X-ray luminosity, and $\chi^2_{\nu}$. We have found correlation between mass fraction of two different dust species and correlation between mass fraction and $\rm{\rm{C_{10\mu m}}}$ for two dust species. Figure~\ref{fig:wave_vs_massfrac_compress} illustrates these results. We can see that  PoAl01 mass fraction is anti-correlated ($r_p$=-0.68) with the MieOl01 mass fraction. It means that porous alumina dominates in the fits as shown in Fig.~\ref{fig:box_plot_final_composition}. Also, from the color bar, we can notice that the larger the PoAl01 mass fraction, the larger the $\rm{\rm{C_{10\mu m}}}$, and the larger the MieOl01 mass fraction, the lower the $\rm{\rm{C_{10\mu m}}}$. This means that porous alumina is contributing to reproduce the $\rm{10\ \mu m}$ silicate feature in cases where it peaks at longer wavelengths, while Mie olivine $0.1\mu$m has the opposite effect.

Taking into account the mass fractions, the particular composition derived from this analysis is $\sim50\%$ PoAl01, $\sim30\%$ MieOl01, $\sim20\%$ Per01, and $\sim0.7\%$ MieOl3. It is worth stressing that the dust composition we are proposing only includes oxygen-rich grains, hence, the mass fractions are based on the total mass of silicates only. In radiative transfer simulations, the amount of carbonaceous dust grains must be also considered within the dust mixture. In this sense, if we assume, as an example, the typical percentages of the ISM: 47\% of carbonaceous grains and 53\% of silicates, then the mass fractions must be multiplied by a factor of 0.53.

\section{Discussion}

In this section, we discuss and compare our results, mainly the dust species that compose our proposed chemical composition in AGN dust.

\subsection{The ``anomalous'' silicate features} \label{sec:the_anomalous}


Several works have reported an anomalous silicate emission feature in the MIR spectra of (typically, but not only) type-1 AGN, characterized by a different spectral shape when compared to that expected by a standard dust mixture as the one observed in the Milky Way ISM \citep[e.g.][]{Sturm2005,Li2008,Smith2010,Lyu2014,Martinez-Paredes2020}. In the spectra of the objects studied in this work, we noticed that most of them show the peak of the $\rm{10\ \mu m}$ silicate feature shifted up to $\rm{\sim0.8\ \mu m}$ towards longer wavelengths with respect to the 9.7 canonical value, and that the peak of the $\rm{18\ \mu m}$ silicate feature is shifted up to $\rm{\sim1.0\ \mu m}$ towards shorter wavelengths with respect to the 18 canonical value. We found that the standard silicates and CDE and Mie olivine with grain sizes between $\sim1$ and $\sim4$ $\mu$m contribute to properly reproducing the $\rm{10\ \mu m}$ silicate emission feature when it shows this shift. This is why these grain sizes are needed to obtain the best fit in almost all objects of our sample (see Table~\ref{tab:best_fits_sample}). This result is in agreement with the finding of \citet{Gonzalez-Martin2023}, who suggest that grain sizes greater than the ones in the range $0.005$–$\rm{0.25 \mu m}$, which is the grain size range commonly used in dusty torus models, are important to reproduce the observed MIR emission. On the other hand, the $\rm{18\ \mu m}$ silicate emission feature shift is well reproduced by $\rm{0.1\ \mu m}$ olivine grains, both CDE and Mie. \citet{Martinez-Paredes2020} find a similar behavior for the silicate features of a sample of 67 type 1 AGN. They find that, on average, the $\rm{10\ \mu m}$ feature peaks at $10.3^{+0.7}_{-0.9}$ $\mu$m and the $\rm{18\ \mu m}$ feature peaks at $17.3^{+0.4}_{-0.7}$ $\mu$m. In our sample, we have 15 objects in common with their sample, and we have 41 type 1 AGN, so that explains the similarity in the characteristics of the spectra of both samples. 

\citet{Li2008} study the deviations of the silicate emission profiles of 3C~273 and NGC~3998 from that of the Galactic ISM. They use a dust model composed of amorphous silicate with olivine composition, and amorphous carbon with optical constants from \citet{Dorschner1995} and \citet{Rouleau1991}, respectively. They consider dust porosity ranging from 0 (compact dust) to 0.9 (fluffy) with a grain size of 0.1~$\mu$m, and blackbody emission. Although they find that combined effects of porosity and temperature suffice by themselves to explain the general trend of broadening the 9.7 $\mu$m silicate emission feature and shifting the peak to longer wavelengths, they do not exclude the effects of other factors such as grain size, shape, and mineralogy. Their results are evidence that dust compositions different from the standard silicates of the ISM can explain the characteristics observed in the MIR spectra of AGN, as our results also show. Another similar work is the one performed by \citet{Smith2010}. They attempt to reproduce the MIR silicate emission of M~81 in order to study its dust composition. They find that the best-fit model is obtained using the optical constants of olivine measured by \citet{Dorschner1995} (the same used in this work). They also tried to fit the MIR silicate emission of M~81 using the dielectric functions of astronomical silicates measured by \citet{DraineLi1984}, but they found that such a model was not as good as the one based on olivine. This is completely in agreement with our results since we found that astronomical silicates provide a much smaller contribution than oxides and amorphous silicates in obtaining good fits in almost all AGN of our sample (see Figure~\ref{fig:bar_plot_dust_species}).

\subsection{Dust composition}

Since a few decades ago, the presence of silicates in AGN has been detected through the spectral features in absorption and emission at $\sim~10$ and $\rm{\sim 18\ \mu m}$ \citep[e.g.][]{Jaffe2004,Hao2005,Siebenmorgen2005,Sturm2005,Hao2007,Nikutta2009,Xie2014,Xie2017}. 
In this work, we have explored if and how the oxides porous alumina and periclase, other than standard silicate grains, can be traced in AGN MIR spectra and found evidence for their presence. This result was also found by \citet{Markwick-Kemper2007} for PG~2112+059 whose MIR spectrum is well reproduced with the effect produced by the two dust species. 
Also, they find that both spherical and non-spherical grains give good results for olivine, Mg-rich olivine, and porous alumina, which is also consistent with our results since our proposed composition includes a mixture of CDE particles (i.e., non-spherical grains) for porous alumina and periclase and Mie particles (i.e., spherical grains) for olivine. 

Our findings have been obtained by implementing a mineralogy model, similar numerical/computational procedures, and dust species used in \citet{Srinivasan2017}, so it is important to compare our results with theirs. They find that the AGN dust of a PG quasars sample is mostly composed of amorphous oxides and silicates, with a small fraction of crystalline silicates. Their analysis suggests that the dust is dominated by porous alumina and olivine with mean mass fractions of $46 \pm 6$ and
$31 \pm 11$ \%, respectively, followed by Mg-rich olivine and periclase with $11 \pm 9$ \% and $9 \pm 1$ \%. Comparing these values with the median mass fractions of our proposed chemical composition, we find that our median mass fraction of porous alumina ($\sim 50$ \%) and olivine ($\sim 30$ \%) are similar. For periclase, we obtained a median mass fraction that is $\sim 10$ \% larger than their value, and finally, for Mg-rich olivine we do not have mass fractions since this dust species is not included in our proposed chemical composition. As for the small fraction of crystalline silicates found by \citet{Srinivasan2017}, we obtained that forsterite and clinoenstatite have negligible mass and flux contributions to the composition, which means that, at least in the objects in our sample, crystalline silicates are not important in the dust chemical composition. Differences between their results and ours could be explained by the differences in the sample since they limited theirs only to quasars with $z<4$ and we studied AGN in general with $z<0.1$. From their sample, there are eight objects in common with ours: I~Zw~1, Mrk~110, Mrk~705, PG~1211+143, SZ~II~010, PG~1351+640, Mrk~841, Mrk~290. We found that, in terms of mass fraction, porous alumina and periclase have a similar importance in the dust composition of these objects in both works. In addition, the objects whose best-fit models (see Table~\ref{tab:best_fits_sample}) have the same composition as the best-fit models from \citet{Srinivasan2017} (viz., Mrk~110, Mrk~705, and Mrk~841), have consistent mass fractions.

\citet{Xie2017} modeled the \emph{Spitzer}/IRS spectra of a sample of 93 AGN in order to calculate the properties of the dust such as grain size and chemical composition. They considered two kinds of dust: carbonaceous dust (graphite and amorphous carbon) and amorphous silicate (the astronomical silicates from \citet{DraineLi1984}, three pyroxene species, and two olivine species). They find that astronomical silicates and graphite provide the best fits for the spectra of 60 out of 93 AGN. This result differs from ours, since we found that astronomical silicates do not have significant importance in the best and good fits (see Figure~\ref{fig:bar_plot_dust_species}) and a composition of pure astronomical silicates can not reproduce the silicate features of our sample. The bottom panel of Figure~\ref{fig:all_residuals} highlights this finding, as it is clear that the residuals of best-fit models obtained adopting the chemical composition of DLSil, are beyond $-10$ and $10\%$ at almost all wavelengths, even reaching discrepancies around $50\%$ in certain spectral windows. The larger residuals are found at $\rm{15 \ \mu m} < \lambda < \rm{20 \ \mu m}$, which means that DLSil can not adequately fit the $\rm{18 \ \mu m}$ silicate feature for most objects of our sample. The same applies for the $\rm{10 \ \mu m}$ silicate feature, but to a lesser degree. In addition, \citet{Xie2017} find that for 31 out of 93 AGN, amorphous olivine combined with graphite can give the best fits and amorphous pyroxene provides the best fit of the spectra for two objects. These results coincide with ours since we obtained that olivine is needed to get the best fit for $\sim40\%$ of our sample and pyroxene (clinoenstatite) has a negligible contribution. Differences in the results can be explained by differences in the sample and assumptions in the dust model since they study quasars and consider two temperature components (cool and warm) for the dust, respectively.

\section{Conclusions}

We have implemented the SED fitting technique on the \emph{Spitzer}/IRS (5-32 $\mu$m) spectra of 49 AGN, and a mineralogy model described by a linear combination of a number of dust species that are meant to properly represent the emission features at $\rm{10}$ and $\rm{18\ \mu m}$ from silicate-based grains that are observed in the MIR spectra of AGN. These dust species consist of oxides, amorphous silicates, and crystalline silicates; also including the standard astronomical silicates that have been used for decades in the literature to calculate dusty torus emission models. They are superposed over a continuum function meant to represent the thermal continuum emission of graphites and silicates. 

In our analysis, one possible source of degeneracy is the choice of this continuum function. We have tested this aspect by using two different prescriptions for this: a power law and a cubic function, concluding that choosing one or the other does not affect our results, as both the $\chi^2_{\nu}$ and AIC values obtained in the best fits using these functions were similar. 

The parameters of each dust species calculated in the fitting models, are the continuum-subtracted extinction efficiencies $Q_{\text{ext,cs}}^{\lambda}$. Since they are calculated from the optical constants for a given grain size, we used $Q_{\text{ext,cs}}^{\lambda}$ for different grain sizes for those dust species for which optical properties are available (olivine, DLSil and MinSil). When  compared to other works adopting a similar approach, this work has the advantage that the largest set of optical properties has been used, with which we try to fit the spectra of a number of objects displaying a variety of characteristics concerning the silicate features. Like this, we hope to get a better insight into the kind of dust which is present in a such extreme environment and provide useful hints to construct better radiative transfer models of the dusty torus.


There is a large number of dust mixtures composed of dust species with different grain sizes and shapes (spheres and ellipsoids) that can be used to obtain a good fit of the MIR spectra of the 49 AGN with silicate emission features studied in this work, leading to the result that large and small olivine grains produce the longer-wavelength shifted $\rm{10\ \mu m}$ and shorter-wavelength shifted $\rm{18\ \mu m}$ silicate emission features, respectively. However, we have found that the dust species that are more appropriate to represent the chemical composition of our AGN sample, is a combination of porous alumina (Al$_2$O$_3$) and periclase (MgO) with grain sizes composed by a continuous distribution of ellipsoids with a characteristic grain size of $\rm{0.1\ \mu m}$, and olivine (MgFeSiO$_4$) with spherical grain sizes of 0.1 and $\rm{3\ \mu m}$. This is justified by the ability of such a mixture to successfully reproduce the observed characteristics of the silicate features in the majority of the MIR spectra of our sample. 

In addition, we have found that dust composed of pure astronomical silicates (those from \cite{DraineLi1984}), which is commonly used by the current dusty torus models in the literature, can not reproduce the silicate features of our sample. The feasibility and effectiveness of our proposed chemical composition for AGN dust will be tested in subsequent RT simulations of the dusty torus and compared with those that commonly assume standard astronomical silicates. It is worthwhile to mention that with the new observations of James Webb Space Telescope (JWST) the surrounding material within AGNs (unresolved with Spitzer) could be resolved and detected in nearby AGN galaxies, so questions about the dust properties can be easily addressed in future works.

\section*{Acknowledgements}

The authors are grateful to the referee for their thorough review of this paper and for the suggestions and comments that helped improving its quality and clarity. We thank to the technical staff from IRyA: Daniel Díaz, Alfonso Ginori and Gilberto Zavala. We thank Daniel Guirado and Anibal Sierra for their willingness to answer the first author's questions about the dust. We thank the Combined Atlas of Sources with Spitzer IRS Spectra (CASSIS) for providing the data to the community. We acknowledge the developers of the Python packages Numpy, Matplotlib, Scipy, and Miepython, which proved very useful for the analysis presented in this work. OURA thanks CONACyT for the Ph.D. scholarship No. 881295. OURA and JF acknowledge financial support from the UNAM- DGAPA-PAPIIT IN110723 grant, México. SS acknowledges support from UNAM-PAPIIT Program IA 104822. CRA acknowledges financial support from the European Union's Horizon 2020 research and innovation programme under Marie Sklodowska-Curie grant agreement No 860744 (BiD4BESt), from the State Research Agency (AEI-MCINN) and from the Spanish MCINN under grants ``Feeding and feedback in active galaxies", with reference PID2019-106027GB-C42, the project ``Quantifying the impact of quasar feedback on galaxy evolution'', with reference EUR2020-112266, funded by MICIN/AEI/10.13039/501100011033 and the European Union NextGenerationEU/PRTR. MS is supported by the Ministry of Science, Technological Development and Innovations of the Republic of Serbia through the contract No.~451-03-9/2023-14/200002.


\section*{Data Availability}

The data underlying this article, such as mass fractions and spectral contributions fractions obtained from the best fits, are available in the article. If the reader is interested in obtaining the spectra, the best-fit models and/or the optical properties of the dust species, a request should be sent to the first author.




\bibliographystyle{mnras}
\bibliography{References} 




\appendix

\section{Details of the fit procedure} \label{append:statistical_procedure}

The Akaike information criterion \citep[AIC,][]{Akaike1998} is one of the statistical estimators used in this work to measure the goodness of the fits. It estimates the quality of each model, relative to each of other models. The AIC has been implemented by \citet{Esparza-Arredondo2021,Martinez-Paredes2021,Gonzalez-Martin2023} to obtain the best fit. The AIC value of the model is defined by

\begin{equation} \label{eq:AIC}
    \text{AIC} = 2k - 2\ln{(P)},
\end{equation}

\noindent where $k$ is the number of free parameters while $P$ is the value of the likelihood function for each model. In this study, the term of the likelihood was defined as following:

\begin{equation} \label{eq:chi2}
   - 2\ln{(P)} \equiv \chi^2 = \mathlarger{\mathlarger{\sum}}_i \left(\frac{\text{data}_i- \text{model}_i}{\text{error}_i}\right)^2,
\end{equation}

\noindent where error$_i$ is the error associated to the $i^{\text{th}}$ datum. Under this definition, the $\chi^2$ of the best fit should approach to the degree of freedom ($\nu$), provided that the uncertainties are Gaussian. Thus, letting be $n$ the number of data points, the reduced $\chi^2$ ($\chi^2_{\nu}$) given by

\begin{equation} \label{eq:chi2_nu}
    \chi^2_{\nu} = \frac{\chi^2}{\nu} = \frac{\chi^2}{n-k},
\end{equation} 

\noindent should be close to unity. In this sense, Equation~\ref{eq:AIC} becomes

\begin{equation} \label{eq:AIC_chi2}
    \text{AIC} = 2k + \chi^2.
\end{equation}

In this work, $\Delta$AIC values have been calculated to compare our computed fits to get the best and good fits (or models). $\Delta$AIC is the relative difference between the best model and each other model in the set. The formula is

\begin{equation} \label{eq:delta_AIC}
    \Delta \text{AIC} = \text{AIC}_i - \text{AIC}_{\text{min}},
\end{equation}

\noindent where AIC$_i$ is the value for the model $i$ and AIC$_{\text{min}}$ is the value for the best model. \citet{Burnham2004} give the following rule of thumb for interpreting $\Delta$AIC values: Models having $\Delta$AIC~$\lid~2$ have substantial support (evidence), those in which $4\lid\Delta$AIC $\lid7$ have considerably less support, and models having ${\rm \Delta AIC >10}$ have essentially no support. As a first step in selecting good quality fits (those defined as good fits and described in Section~\ref{sec:results}), we eliminate all models with ${\rm \Delta AIC >10}$.

In addition to the AIC, to test the goodness of our fits, we have used the $\chi^2_{\nu}$ value calculated with Equation~(\ref{eq:chi2_nu}) and the residuals calculated by following

\begin{equation}
    R = 100 \cdot \frac{\text{data}_i- \text{model}_i}{\text{model}_i},
\end{equation}

\noindent and its error with

\begin{equation}
    R_{\text{error}} = 100 \cdot \frac{\text{error}_i}{\text{model}_i},
\end{equation}

\section{Species of our dust mixture} \label{append:dust_species}

In the following we investigate the nature of porous alumina, periclase, and olivine, their possible formation mechanisms, astrophysical environments where the presence of them have been detected, and their previous use in the analysis and modeling of observed SED.

\subsection{Porous alumina}
According to the thermodynamic condensation sequence for gas of solar composition, the silicate condensation sequence starts with the formation of porous alumina (Al$_2$O$_3$). This occurs in a temperature range 1300 - 1760 K \citep{Tielens1998,Tielens1990}. This material has been found in circumstellar shells \citep{Tielens1998} and stellar winds of asymptotic giant branch (AGB) stars \citep[e.g.][]{Tielens2022,Ventura2020,Blommaert2006}. In particular, \citeauthor{Blommaert2006} report that the SED of the 70\% of their observed  sample of low mass-loss rate AGB stars can be reproduced with pure porous alumina. In addition, \citet{Elvis2002} argue that the formation of porous alumina and other dust species can take place through dust condensation in a chemically enriched medium with a sufficiently low temperature, and a large enough density. The amount of dust, chemical composition, and grain size is given by the duration of the favorable conditions. This scenario of dust formation is called ``dust formation window''. They accept dust formation around AGB shells as a fact and use the similarity in physical conditions with AGN tori to imply that the latter environment is also ideal for dust formation. Interestingly enough, these temperature and density conditions are very similar to those that can be found in the AGN environment \citep{Barvainis1987}.

\subsection{Periclase}

According to \citet{Ferrarotti2003}, the formation of silicates (e.g. olivine, Mg-rich olivine, forsterite, clinoenstatite) in stellar winds is related to the formation of periclase (MgO). The dust formation process of silicates is not completed until the gas temperature is below the stability limit of periclase ($\sim 1050$ K) and a significant fraction of the magnesium in the gas phase is left at this temperature to form periclase. \citet{Nishida2022} construct a galaxy SED model including the dust evolution in galaxies which considers dust species within which MgO is included. They assume that dust is originated by evolutionary processes that occur in AGB stars and type-II supernovae (SNe II). Our results suggest that periclase is present in AGN as well.

\subsection{Olivine}

Olivine (along with pyroxene) constitutes the bulk of the mass of the ISM silicates \citep{Zhukovska2008}. The latter work studies the dust evolution in the solar neighborhood and in the Milky Way disk. They find that, in this environment, silicates form mostly in molecular clouds, followed by AGB stars and SNe II. In contrast, in the Galactic disk, the majority of silicates come from the massive AGB stars of supersolar metallicity, with short lifetimes. Olivine is commonly used to model the dust composition of AGB stars. However, \citet{Marini2023} explore optical constants of \citet{DraineLi1984} and \citet{Dorschner1995} for olivine and find a poor agreement between the models and data for a sample of AGB stars. \citet{Kozasa1989} study the dust formation in the ejecta of SNe. They mention that dust grains can nucleate and grow only in the metal-rich cooling gas. In this sense, \citet{Nozawa2003,Nozawa2007} investigate the dust formation in Population III SNe. They find that Al$_2$O$_3$, MgO, olivines, and a variety of other species, condense in different layers in the unmixed ejecta with the original onion-like structure. Recently, using spectra from VLTI/MATTISE, \citet{GamezRosas2022} found the presence of olivine in the AGN galaxy NGC~1068 that is producing the silicate absorption feature at $\rm{10 \mu m}$. 


\bsp	
\label{lastpage}
\end{document}